%% file: main.tex
\documentclass[format=acmsmall, review=false, screen=true]{acmart}

\usepackage{booktabs} 

\acmVolume{1}
\acmNumber{1}
\acmArticle{1}
\acmYear{201X}
\acmMonth{1}

\acmDOI{0000001.0000001}

\usepackage{epstopdf}
\usepackage{textcomp} 
\usepackage{adjustbox} 
\usepackage{textcase}
\usepackage{amsfonts}
\usepackage{graphicx}
\RequirePackage{xspace}
\RequirePackage{graphics}
\usepackage{xcolor}
\RequirePackage{textcomp}
\usepackage{mathrsfs}
\usepackage{lipsum}
\usepackage{listings}
\usepackage{multirow}
\usepackage[english]{babel}
\usepackage{array}
\usepackage{pifont} 
\usepackage{stmaryrd}
\usepackage{footnote}
\usepackage{makecell}
\usepackage{xparse}

\usepackage{courier}

\usepackage{amssymb}
\usepackage{graphicx}
\usepackage{amsmath}
\usepackage{amsfonts}
\RequirePackage{xspace}
\RequirePackage{graphics}
\usepackage{xcolor}
\RequirePackage{textcomp}
\usepackage{keyval}
\usepackage{xspace}
\usepackage{mathrsfs}
\usepackage{paralist}
\usepackage{lipsum}
\usepackage{listings}
\usepackage{multirow}
\usepackage{array}
\usepackage{pifont} 
\usepackage{stmaryrd}
\usepackage{adjustbox} 
\usepackage{textcase}
\usepackage{makecell}
\usepackage[inline]{enumitem} 
\usepackage{flushend} 

\usepackage{standalone}
\newif\ifbuildtikz

\buildtikztrue

\RequirePackage{tikz}
\usetikzlibrary{arrows,automata,shapes,calc,through,decorations.pathmorphing,decorations.fractals,chains,shapes.multipart}

\definecolor{javared}{rgb}{0.6,0,0} 
\definecolor{javagreen}{rgb}{0.25,0.5,0.35} 
\definecolor{javapurple}{rgb}{0.5,0,0.35} 
\definecolor{javadocblue}{rgb}{0.25,0.35,0.75} 

\usepackage{arydshln} 

\usepackage{circuitikz}

\hypersetup{
  colorlinks				=true,
  citecolor					={purple},
  linkcolor 				={blue},
  bookmarksopen			=true,
  bookmarksnumbered	=true
}

\usepackage{etoolbox}

\usepackage[T1]{fontenc}
\usepackage[latin9]{inputenc}
\usepackage{float}
\usepackage{amsmath}
\usepackage{amssymb}

\usepackage{subfig}

\makeatletter
\DeclareRobustCommand*\cal{\@fontswitch\relax\mathcal}

\floatstyle{ruled}
\newfloat{algorithm}{tbp}{loa}
\providecommand{\algorithmname}{Algorithm}
\floatname{algorithm}{\protect\algorithmname}

\date{}

\makeatother

\usepackage{listings}

\def\papertype{paper\xspace}
\newtoggle{extended_version}
\togglefalse{extended_version}

\include{prelude}

\def\titlename{Cyber-Physical Specification Mismatches}

\begin{document}

\title{\titlename}


\author{Luan V. Nguyen}
\affiliation{University of Texas at Arlington}
\author{Khaza Anuarul Hoque}
\affiliation{University of Oxford}
\author{Stanley Bak}
\affiliation{Air Force Research Laboratory}
\author{Steven Drager}
\affiliation{Air Force Research Laboratory}
\author{Taylor T. Johnson}
\affiliation{Vanderbilt University}


\begin{abstract}
\input{abstract.tex}
%
\end{abstract}



%

\maketitle
\renewcommand{\shortauthors}{L. V. Nguyen et al.}{\titlename}

\input{intro}


\input{design_reuse}

\input{model}

\input{specification_mismatch}


\input{tool}

\input{experiment}
\input{discussion}

\input{conclusion}
\vspace{-0.5em}
\begin{acks}
%
%
The material presented in this paper is based upon work supported by
the National Science Foundation (NSF) under grant numbers CNS 1464311,
CNS 1713253, EPCN 1509804, SHF 1527398, and SHF 1736323, the Air Force
Research Laboratory (AFRL) through the AFRL's Visiting Faculty
Research Program (VFRP) under contract number FA8750-13-2-0115, as
well as contract numbers FA8750-15-1-0105, and FA8650-12-3-7255 via
subcontract number WBSC 7255 SOI VU 0001, and the Air Force Office of
Scientific Research (AFOSR) through AFOSR's Summer Faculty Fellowship
Program (SFFP) under contract number FA9550-15-F-0001, as well as
under contract numbers FA9550-15-1-0258 and FA9550-16-1-0246. The U.S.
government is authorized to reproduce and distribute reprints for
Governmental purposes notwithstanding any copyright notation thereon.
Any opinions, findings, and conclusions or recommendations expressed
in this publication are those of the authors and do not necessarily
reflect the views of AFRL, AFOSR, or NSF.
\end{acks}
\vspace{-0.25em}
\renewcommand{\baselinestretch}{0.94}
\bibliographystyle{ACM-Reference-Format}
\addcontentsline{toc}{section}{\refname}\bibliography{master,luan,bak}

\end{document}

%% file: prelude.tex
\newcommand{\todo}[1]{}

\newcommand{\num}[1]{\relax\ifmmode \mathbb #1\else $\mathbb #1$\fi}
\newcommand{\nnnum}[1]{\relax\ifmmode 
  {\mathbb #1}_{\geq 0} \else ${\mathbb #1}_{\geq 0}$
  \fi}
\newcommand{\npnum}[1]{\relax\ifmmode 
  {\mathbb #1}_{\leq 0} \else ${\mathbb #1}_{\leq 0}$
  \fi}
\newcommand{\pnum}[1]{\relax\ifmmode 
  {\mathbb #1}_{> 0} \else ${\mathbb #1}_{> 0}$
  \fi}
\newcommand{\nnum}[1]{\relax\ifmmode 
  {\mathbb #1}_{< 0} \else ${\mathbb #1}_{< 0}$
  \fi}
\newcommand{\plnum}[1]{\relax\ifmmode 
  {\mathbb #1}_{+} \else ${\mathbb #1}_{+}$
  \fi}
\newcommand{\nenum}[1]{\relax\ifmmode 
  {\mathbb #1}_{-} \else ${\mathbb #1}_{-}$
  \fi}

\newcommand{\extb}[1]{\relax\ifmmode {\sf ExtBeh}_{#1} \else ${\sf ExtBeh}_{#1}$\fi} 
\newcommand{\tdists}[1]{\relax\ifmmode {\sf Tdists}_{#1} \else ${\sf Tdists}_{#1}$\fi} 

\newcommand{\exec}[1]{\relax\ifmmode {\sf Execs}_{#1} \else ${\sf Exec}_{#1}$\fi} 
\newcommand{\execf}[1]{\relax\ifmmode {\sf Execs}^*_{#1} \else ${\sf Exec}^*_{#1}$\fi} 
\newcommand{\execi}[1]{\relax\ifmmode {\sf Execs}^\omega_{#1} \else ${\sf Exec}^\omega_{#1}$\fi} 

\newcommand{\ctrace}[1]{\relax\ifmmode {\sf Ctraces}_{#1} \else ${\sf Ctraces}_{#1}$\fi} 

\newcommand{\trace}[1]{\relax\ifmmode {\sf Traces}_{#1} \else ${\sf Traces}_{#1}$\fi} 
\newcommand{\tracef}[1]{\relax\ifmmode {\sf Traces}^*_{#1} \else ${\sf Traces}^*_{#1}$\fi} 
\newcommand{\tracei}[1]{\relax\ifmmode {\sf Traces}^\omega_{#1} \else ${\sf Traces}^\omega_{#1}$\fi} 

\newcommand{\frag}[1]{\relax\ifmmode {\sf Frags}_{#1} \else ${\sf Frags}_{#1}$\fi} 
\newcommand{\fragf}[1]{\relax\ifmmode {\sf Frags}^*_{#1} \else ${\sf Frags}^*_{#1}$\fi} 
\newcommand{\fragi}[1]{\relax\ifmmode {\sf Frags}^\omega_{#1} \else ${\sf Frags}^\omega_{#1}$\fi} 

\newcommand{\reach}[1]{\relax\ifmmode {\sf Reach}_{#1} \else ${\sf Reach}_{#1}$\fi}

\newcommand{\type}[1]{\ms{type{(#1)}}}

\def\A{{\cal A}} 
\def\E{{\cal E}} 
\def\I{{\cal I}} 
\def\Q{{\cal Q}} 
\def\R{{\cal R}} 
\def\T{{\cal T}} 
\def\V{{\cal V}} 
\def\U{{\cal U}} 

\newcommand{\col}[1]{\relax\ifmmode \mathscr #1\else $\mathscr #1$\fi}

\definecolor{HIOAcolor}{rgb}{0.776,0.22,0.07}

\newcommand{\SC}[2]{\relax\ifmmode {\tt Scount}(#1,#2) \else ${\tt Scount}(#1,#2)$\fi} 
\newcommand{\SCM}[2]{\relax\ifmmode {\tt Smin}(#1,#2) \else ${\tt Smin}(#1,#2)$\fi} 
\newcommand{\Aut}[1]{\relax\ifmmode {\tt Aut}(#1) \else ${\tt Aut}(#1)$\fi}

\newcommand{\act}[1]{{\operatorname{\mathsf{#1}}}}

\newcommand{\Var}{{\bf var}}

\newcommand{\deq}{\mathrel{\stackrel{\scriptscriptstyle\Delta}{=}}}

\newcommand{\seclabel}[1]{\label{sec:#1}}
\newcommand{\secref}[1]{Section~\ref{sec:#1}}

\newcommand{\figlabel}[1]{\label{fig:#1}}
\newcommand{\figref}[1]{Figure~\ref{fig:#1}}

\newcommand{\tablabel}[1]{\label{tab:#1}}
\newcommand{\tabref}[1]{Table~\ref{tab:#1}}

\newcommand{\deflabel}[1]{\label{def:#1}}
\newcommand{\defref}[1]{Definition~\ref{def:#1}}

\newcommand{\eqlabel}[1]{\label{eq:#1}}
\renewcommand{\eqref}[1]{Equation~\ref{eq:#1}}

\newcommand{\remove}[1]{}
\newcommand{\salg}[1]{\relax\ifmmode {\mathcal F}_{#1}\else ${\mathcal F}_{#1}$\fi} 
\newcommand{\msp}[1]{\relax\ifmmode (#1, \salg{#1}) \else $(#1, \salg{#1})$\fi} 
\newcommand{\msprod}[2]{\relax\ifmmode ( #1 \times #2, \salg{#1} \otimes \salg{#2}) \else $(#1 \times #2, \salg{#1} \otimes \salg{#2})$\fi} 
\newcommand{\dist}[1]{\relax\ifmmode {\mathcal P}\msp{#1}
  \else ${\mathcal P}\msp{#1}$\fi} 
\newcommand{\subdist}[1]{\relax\ifmmode {\mathcal S}{\mathcal P}\msp{#1} 
  \else ${\mathcal S}{\mathcal P}\msp{#1}$\fi} 
\newcommand{\disc}[1]{\relax\ifmmode {\sf Disc}(#1)
  \else ${\sf Disc}(#1)$\fi} 

\newcommand{\Trajeq}{\relax\ifmmode {\mathcal R}_\T \else ${\mathcal R}_\T$\fi} 
\newcommand{\Acteq}{\relax\ifmmode {\mathcal R}_A \else ${\mathcal R}_A$\fi} 
\newcommand{\noop}{\relax\ifmmode \lambda \else $\lambda$\fi} 
\newcommand{\close}[1]{\relax\ifmmode \overline{#1} \else $\overline{#1}$\fi}

\newcommand{\tup}[1]
           {
             \relax\ifmmode
             \langle #1 \rangle
             \else $\langle$ #1 $\rangle$ \fi
           }

\newcommand{\lit}[1]{ \relax\ifmmode
                \mathord{\mathcode`\-="702D\sf #1\mathcode`\-="2200}
                \else {\it #1} \fi }

\newcommand{\figuresize}{\scriptsize}

\lstdefinelanguage{ioa}{
  basicstyle=\figuresize,
  keywordstyle=\bf \figuresize,
  identifierstyle=\it \figuresize,
  emphstyle=\tt \figuresize,
  mathescape=true,
  tabsize=20,
  sensitive=false,
  columns=fullflexible,
  keepspaces=false,
  flexiblecolumns=true,
  basewidth=0.05em,
  escapeinside={(*@}{@*)},
  moredelim=[il][\rm]{//},
  moredelim=[is][\sf \figuresize]{!}{!},
  moredelim=[is][\bf \figuresize]{*}{*},
  keywords={automaton,and, 
  	 choose,const,continue, components,
  	 discrete, do,
  	 eff, Eff, external,else, elseif, evolve, end,
  	 fi,for, forward, from,
  	 hidden,
  	 in,input,internal,if,invariant, initially, imports,
     let,
     or, output, operators, od, of,
     pre, Pre,
     return,
     such,satisfies, stop, signature, simulation, 
     trajectories,trajdef, transitions, that,then, type, types, to, tasks,
     variables, vocabulary, 
     when,where, with,while},
  emph={set, seq, tuple, map, array, enumeration},   
   literate=
        {(}{{$($}}1
        {)}{{$)$}}1
        {\\in}{{$\in\ $}}1
        {\\preceq}{{$\preceq\ $}}1
        {\\subset}{{$\subset\ $}}1
        {\\subseteq}{{$\subseteq\ $}}1
        {\\supset}{{$\supset\ $}}1
        {\\supseteq}{{$\supseteq\ $}}1
        {\\forall}{{$\forall$}}1
        {\\le}{{$\le\ $}}1
        {\\ge}{{$\ge\ $}}1
        {\\gets}{{$\gets\ $}}1
        {\\cup}{{$\cup\ $}}1
        {\\cap}{{$\cap\ $}}1
        {\\langle}{{$\langle$}}1
        {\\rangle}{{$\rangle$}}1
        {\\exists}{{$\exists\ $}}1
        {\\bot}{{$\bot$}}1
        {\\rip}{{$\rip$}}1
        {\\emptyset}{{$\emptyset$}}1
        {\\notin}{{$\notin\ $}}1
        {\\not\\exists}{{$\not\exists\ $}}1
        {\\ne}{{$\ne\ $}}1
        {\\to}{{$\to\ $}}1
        {\\implies}{{$\implies\ $}}1
        {<}{{$<\ $}}1
        {>}{{$>\ $}}1
        {=}{{$=\ $}}1
        {~}{{$\neg\ $}}1
        {|}{{$\mid$}}1
        {'}{{$^\prime$}}1
        {\\A}{{$\forall\ $}}1
        {\\E}{{$\exists\ $}}1
        {\\nE}{{$\nexists\ $}}1
        {\\/}{{$\vee\,$}}1
        {\\vee}{{$\vee\,$}}1
        {/\\}{{$\wedge\,$}}1
        {\\wedge}{{$\wedge\,$}}1
        {=>}{{$\Rightarrow\ $}}1
        {->}{{$\rightarrow\ $}}1
        {<=}{{$\Leftarrow\ $}}1
        {<-}{{$\leftarrow\ $}}1
        {~=}{{$\neq\ $}}1
        {\\U}{{$\cup\ $}}1
        {\\I}{{$\cap\ $}}1
        {|-}{{$\vdash\ $}}1
        {-|}{{$\dashv\ $}}1
        {<<}{{$\ll\ $}}2
        {>>}{{$\gg\ $}}2
        {||}{{$\|$}}1
        {[}{{$[$}}1
        {]}{{$\,]$}}1
        {[[}{{$\langle$}}1
        {]]]}{{$]\rangle$}}1
        {]]}{{$\rangle$}}1
        {<=>}{{$\Leftrightarrow\ $}}2
        {<->}{{$\leftrightarrow\ $}}2
        {(+)}{{$\oplus\ $}}1
        {(-)}{{$\ominus\ $}}1
        {_i}{{$_{i}$}}1
        {_j}{{$_{j}$}}1
        {_{i,j}}{{$_{i,j}$}}3
        {_{j,i}}{{$_{j,i}$}}3
        {_0}{{$_0$}}1
        {_1}{{$_1$}}1
        {_2}{{$_2$}}1
        {_n}{{$_n$}}1
        {_p}{{$_p$}}1
        {_k}{{$_n$}}1
        {-}{{$\ms{-}$}}1
        {@}{{}}0
        {\\delta}{{$\delta$}}1
        {\\R}{{$\R$}}1
        {\\Rplus}{{$\Rplus$}}1
        {\\N}{{$\N$}}1
        {\\times}{{$\times\ $}}1
        {\\tau}{{$\tau$}}1
        {\\alpha}{{$\alpha$}}1
        {\\beta}{{$\beta$}}1
        {\\gamma}{{$\gamma$}}1
        {\\ell}{{$\ell\ $}}1
        {--}{{$-\ $}}1
        {\\TT}{{\hspace{1.5em}}}3        
      }

\lstdefinelanguage{ioaNums}[]{ioa}
{
  numbers=left,
  numberstyle=\tiny,
  stepnumber=2,
  numbersep=4pt
}

\lstdefinelanguage{ioaNumsRight}[]{ioa}
{
  numbers=right,
  numberstyle=\tiny,
  stepnumber=2,
  numbersep=4pt
}

\lstnewenvironment{IOA}%
  {\lstset{language=IOA}}
  {}

\lstnewenvironment{IOANums}%
  {
  \if@firstcolumn
    \lstset{language=IOA, numbers=left, firstnumber=auto}
  \else
    \lstset{language=IOA, numbers=right, firstnumber=auto}
  \fi
  }
  {}

\lstnewenvironment{IOANumsRight}%
  {
    \lstset{language=IOA, numbers=right, firstnumber=auto}
  }
  {}

\newcommand{\linefigioa}[9]{

}

\lstdefinelanguage{ioaLang}{%
  basicstyle=\ttfamily\small,
  keywordstyle=\rmfamily\bfseries\small,
  identifierstyle=\small,
  keywords={assumes,automaton,axioms,backward,bounds,by,case,choose,components,const,d,det,discrete,do,eff,else,elseif,ensuring,enumeration,evolve,fi,fire,follow,for,forward,from,hidden,if,in,%
    input,initially,internal,invariant,let, local,od,of,output,pre,schedule,signature,so,%
    simulation,states,variables, tasks, stop,tasks,that,then,to,trajdef,trajectory,trajectories,transitions,tuple,type,union,urgent,uses,when,where,while,yield},
  literate=
        {\\in}{{$\in$}}1
        {\\preceq}{{$\preceq$}}1
        {\\subset}{{$\subset$}}1
        {\\subseteq}{{$\subseteq$}}1
        {\\supset}{{$\supset$}}1
        {\\supseteq}{{$\supseteq$}}1
        {\\rho}{{$\rho$}}1
        {\\infty}{{$\infty$}}1
        {<}{{$<$}}1
        {>}{{$>$}}1
        {=}{{$=$}}1
        {~}{{$\neg$}}1 
        {|}{{$\mid$}}1
        {'}{{$^\prime$}}1
        {\\A}{{$\forall$}}1 {\\E}{{$\exists$}}1
        {\\/}{{$\vee$}}1 {/\\}{{$\wedge$}}1 
        {=>}{{$\Rightarrow$}}1 
        {->}{{$\rightarrow$}}1 
        {<=}{{$\leq$}}1 {>=}{{$\geq$}}1 {~=}{{$\neq$}}1
        {\\U}{{$\cup$}}1 {\\I}{{$\cap$}}1
        {|-}{{$\vdash$}}1 {-|}{{$\dashv$}}1
        {<<}{{$\ll$}}2 {>>}{{$\gg$}}2
        {||}{{$\|$}}1
        {<=>}{{$\Leftrightarrow$}}2 
        {<->}{{$\leftrightarrow$}}2
        {(+)}{{$\oplus$}}1
        {(-)}{{$\ominus$}}1
}

\lstdefinelanguage{bigIOALang}{%
  basicstyle=\ttfamily,
  keywordstyle=\rmfamily\bfseries,
  identifierstyle=,
  keywords={assumes,automaton,axioms,backward,by,case,choose,components,const,%
    d,det,discrete,do,eff,else,elseif,ensuring,enumeration,evolve,fi,for,forward,from,hidden,if,in%
    input,initially,internal,invariant,local,od,of,output,pre,schedule,signature,so,%
    tasks, simulation,states,stop,tasks,that,then,to,trajdef,trajectories,transitions,tuple,type,union,urgent,uses,when,where,yield},
  literate=
        {\\in}{{$\in$}}1
        {\\preceq}{{$\preceq$}}1
        {\\subset}{{$\subset$}}1
        {\\subseteq}{{$\subseteq$}}1
        {\\supset}{{$\supset$}}1
        {\\supseteq}{{$\supseteq$}}1
        {<}{{$<$}}1
        {>}{{$>$}}1
        {=}{{$=$}}1
        {~}{{$\neg$}}1 
        {|}{{$\mid$}}1
        {'}{{$^\prime$}}1
        {\\A}{{$\forall$}}1 {\\E}{{$\exists$}}1
        {\\/}{{$\vee$}}1 {/\\}{{$\wedge$}}1 
        {=>}{{$\Rightarrow$}}1 
        {->}{{$\rightarrow$}}1 
        {<=}{{$\leq$}}1 {>=}{{$\geq$}}1 {~=}{{$\neq$}}1
        {\\U}{{$\cup$}}1 {\\I}{{$\cap$}}1
        {|-}{{$\vdash$}}1 {-|}{{$\dashv$}}1
        {<<}{{$\ll$}}2 {>>}{{$\gg$}}2
        {||}{{$\|$}}1
        {<=>}{{$\Leftrightarrow$}}2 
        {<->}{{$\leftrightarrow$}}2
        {(+)}{{$\oplus$}}1
        {(-)}{{$\ominus$}}1
}

\lstnewenvironment{BigIOA}%
  {\lstset{language=bigIOALang,basicstyle=\ttfamily}
   \csname lst@SetFirstLabel\endcsname}
  {\csname lst@SaveFirstLabel\endcsname\vspace{-4pt}\noindent}

\lstnewenvironment{SmallIOA}%
  {\lstset{language=ioaLang,basicstyle=\ttfamily\scriptsize}
   \csname lst@SetFirstLabel\endcsname}
  {\csname lst@SaveFirstLabel\endcsname\noindent}

\newlength{\bracklen}

\newcommand{\tri}[3]{\ensuremath{\mathit{#1}^\mathit{#2}_\mathit{#3}}}

\newcommand{\sugLocalVars}[2]{\ifthenelse{\equal{}{#2}}%
                             {\tri{localVars}{#1}{desug}}%
                             {\tri{localVars}{#1}{#2,desug}}}
\newcommand{\sugVars}[2]{\ifthenelse{\equal{}{#2}}%
                        {\tri{vars}{#1}{desug}}%
                        {\tri{vars}{#1}{#2,desug}}}

\newenvironment{subSyntax}{\begin{array}{l}}{\end{array}}

\newcommand{\ms}[1]{\ifmmode%
\mathord{\mathcode`-="702D\it #1\mathcode`\-="2200}\else%
$\mathord{\mathcode`-="702D\it #1\mathcode`\-="2200}$\fi}

\def\A{{\cal A}} 
\def\T{{\cal T}} 

\lstdefinelanguage{pvs}{
  basicstyle=\tt \figuresize,
  keywordstyle=\sc \figuresize,
  identifierstyle=\it \figuresize,
  emphstyle=\tt \figuresize,
  mathescape=true,
  tabsize=20,
  sensitive=false,
  columns=fullflexible,
  keepspaces=false,
  flexiblecolumns=true,
  basewidth=0.05em,
  moredelim=[il][\rm]{//},
  moredelim=[is][\sf \figuresize]{!}{!},
  moredelim=[is][\bf \figuresize]{*}{*},
  keywords={and, 
  	 begin,
  	 cases, const,
  	 do,
  	 external, else, exists, end, endcases, endif,
  	 fi,for, forall, from,
  	 hidden,
  	 in, if, importing,
     let, lambda, lemma,
     measure, 
     not,
     or, of,
     return, recursive,
     stop, 
     theory, that,then, type, types, type+, to, theorem,
     var,
     with,while},
  emph={nat, setof, sequence, eq, tuple, map, array, enumeration, bool, real, exp, nnreal, posreal},   
   literate=
        {(}{{$($}}1
        {)}{{$)$}}1
        {\\in}{{$\in\ $}}1
        {\\mapsto}{{$\rightarrow\ $}}1
        {\\preceq}{{$\preceq\ $}}1
        {\\subset}{{$\subset\ $}}1
        {\\subseteq}{{$\subseteq\ $}}1
        {\\supset}{{$\supset\ $}}1
        {\\supseteq}{{$\supseteq\ $}}1
        {\\forall}{{$\forall$}}1
        {\\le}{{$\le\ $}}1
        {\\ge}{{$\ge\ $}}1
        {\\gets}{{$\gets\ $}}1
        {\\cup}{{$\cup\ $}}1
        {\\cap}{{$\cap\ $}}1
        {\\langle}{{$\langle$}}1
        {\\rangle}{{$\rangle$}}1
        {\\exists}{{$\exists\ $}}1
        {\\bot}{{$\bot$}}1
        {\\rip}{{$\rip$}}1
        {\\emptyset}{{$\emptyset$}}1
        {\\notin}{{$\notin\ $}}1
        {\\not\\exists}{{$\not\exists\ $}}1
        {\\ne}{{$\ne\ $}}1
        {\\to}{{$\to\ $}}1
        {\\implies}{{$\implies\ $}}1
        {<}{{$<\ $}}1
        {>}{{$>\ $}}1
        {=}{{$=\ $}}1
        {~}{{$\neg\ $}}1
        {|}{{$\mid$}}1
        {'}{{$^\prime$}}1
        {\\A}{{$\forall\ $}}1
        {\\E}{{$\exists\ $}}1
        {\\/}{{$\vee\,$}}1
        {\\vee}{{$\vee\,$}}1
        {/\\}{{$\wedge\,$}}1
        {\\wedge}{{$\wedge\,$}}1
        {->}{{$\rightarrow\ $}}1
        {=>}{{$\Rightarrow\ $}}1
        {->}{{$\rightarrow\ $}}1
        {<=}{{$\Leftarrow\ $}}1
        {<-}{{$\leftarrow\ $}}1
        {~=}{{$\neq\ $}}1
        {\\U}{{$\cup\ $}}1
        {\\I}{{$\cap\ $}}1
        {|-}{{$\vdash\ $}}1
        {-|}{{$\dashv\ $}}1
        {<<}{{$\ll\ $}}2
        {>>}{{$\gg\ $}}2
        {||}{{$\|$}}1
        {[}{{$[$}}1
        {]}{{$\,]$}}1
        {[[}{{$\langle$}}1
        {]]]}{{$]\rangle$}}1
        {]]}{{$\rangle$}}1
        {<=>}{{$\Leftrightarrow\ $}}2
        {<->}{{$\leftrightarrow\ $}}2
        {(+)}{{$\oplus\ $}}1
        {(-)}{{$\ominus\ $}}1
        {_i}{{$_{i}$}}1
        {_j}{{$_{j}$}}1
        {_{i,j}}{{$_{i,j}$}}3
        {_{j,i}}{{$_{j,i}$}}3
        {_0}{{$_0$}}1
        {_1}{{$_1$}}1
        {_2}{{$_2$}}1
        {_n}{{$_n$}}1
        {_p}{{$_p$}}1
        {_k}{{$_n$}}1
        {-}{{$\ms{-}$}}1
        {@}{{}}0
        {\\delta}{{$\delta$}}1
        {\\R}{{$\R$}}1
        {\\Rplus}{{$\Rplus$}}1
        {\\N}{{$\N$}}1
        {\\times}{{$\times\ $}}1
        {\\tau}{{$\tau$}}1
        {\\alpha}{{$\alpha$}}1
        {\\beta}{{$\beta$}}1
        {\\gamma}{{$\gamma$}}1
        {\\ell}{{$\ell\ $}}1
        {--}{{$-\ $}}1
        {\\TT}{{\hspace{1.5em}}}3        
      }

\lstdefinelanguage{BigPVS}{
  basicstyle=\tt,
  keywordstyle=\sc,
  identifierstyle=\it,
  emphstyle=\tt ,
  mathescape=true,
  tabsize=20,
  sensitive=false,
  columns=fullflexible,
  keepspaces=false,
  flexiblecolumns=true,
  basewidth=0.05em,
  moredelim=[il][\rm]{//},
  moredelim=[is][\sf \figuresize]{!}{!},
  moredelim=[is][\bf \figuresize]{*}{*},
  keywords={and, 
  	 begin,
  	 cases, const,
  	 do, datatype,
  	 external, else, exists, end, endif, endcases,
  	 fi,for, forall, from,
  	 hidden,
  	 in, if, importing,
     let, lambda, lemma,
     measure,
     not,
     or, of,
     return, recursive,
     stop, 
     theory, that,then, type, types, type+, to, theorem,
     var,
     with,while},
  emph={nat, setof, sequence, eq, tuple, map, array, first, rest, add, enumeration, bool, real, posreal, nnreal},   
   literate=
        {(}{{$($}}1
        {)}{{$)$}}1
        {\\in}{{$\in\ $}}1
        {\\mapsto}{{$\rightarrow\ $}}1
        {\\preceq}{{$\preceq\ $}}1
        {\\subset}{{$\subset\ $}}1
        {\\subseteq}{{$\subseteq\ $}}1
        {\\supset}{{$\supset\ $}}1
        {\\supseteq}{{$\supseteq\ $}}1
        {\\forall}{{$\forall$}}1
        {\\le}{{$\le\ $}}1
        {\\ge}{{$\ge\ $}}1
        {\\gets}{{$\gets\ $}}1
        {\\cup}{{$\cup\ $}}1
        {\\cap}{{$\cap\ $}}1
        {\\langle}{{$\langle$}}1
        {\\rangle}{{$\rangle$}}1
        {\\exists}{{$\exists\ $}}1
        {\\bot}{{$\bot$}}1
        {\\rip}{{$\rip$}}1
        {\\emptyset}{{$\emptyset$}}1
        {\\notin}{{$\notin\ $}}1
        {\\not\\exists}{{$\not\exists\ $}}1
        {\\ne}{{$\ne\ $}}1
        {\\to}{{$\to\ $}}1
        {\\implies}{{$\implies\ $}}1
        {<}{{$<\ $}}1
        {>}{{$>\ $}}1
        {=}{{$=\ $}}1
        {~}{{$\neg\ $}}1
        {|}{{$\mid$}}1
        {'}{{$^\prime$}}1
        {\\A}{{$\forall\ $}}1
        {\\E}{{$\exists\ $}}1
        {\\/}{{$\vee\,$}}1
        {\\vee}{{$\vee\,$}}1
        {/\\}{{$\wedge\,$}}1
        {\\wedge}{{$\wedge\,$}}1
        {->}{{$\rightarrow\ $}}1
        {=>}{{$\Rightarrow\ $}}1
        {->}{{$\rightarrow\ $}}1
        {<=}{{$\Leftarrow\ $}}1
        {<-}{{$\leftarrow\ $}}1
        {~=}{{$\neq\ $}}1
        {\\U}{{$\cup\ $}}1
        {\\I}{{$\cap\ $}}1
        {|-}{{$\vdash\ $}}1
        {-|}{{$\dashv\ $}}1
        {<<}{{$\ll\ $}}2
        {>>}{{$\gg\ $}}2
        {||}{{$\|$}}1
        {[}{{$[$}}1
        {]}{{$\,]$}}1
        {[[}{{$\langle$}}1
        {]]]}{{$]\rangle$}}1
        {]]}{{$\rangle$}}1
        {<=>}{{$\Leftrightarrow\ $}}2
        {<->}{{$\leftrightarrow\ $}}2
        {(+)}{{$\oplus\ $}}1
        {(-)}{{$\ominus\ $}}1
        {_i}{{$_{i}$}}1
        {_j}{{$_{j}$}}1
        {_{i,j}}{{$_{i,j}$}}3
        {_{j,i}}{{$_{j,i}$}}3
        {_0}{{$_0$}}1
        {_1}{{$_1$}}1
        {_2}{{$_2$}}1
        {_n}{{$_n$}}1
        {_p}{{$_p$}}1
        {_k}{{$_n$}}1
        {-}{{$\ms{-}$}}1
        {@}{{}}0
        {\\delta}{{$\delta$}}1
        {\\R}{{$\R$}}1
        {\\Rplus}{{$\Rplus$}}1
        {\\N}{{$\N$}}1
        {\\times}{{$\times\ $}}1
        {\\tau}{{$\tau$}}1
        {\\alpha}{{$\alpha$}}1
        {\\beta}{{$\beta$}}1
        {\\gamma}{{$\gamma$}}1
        {\\ell}{{$\ell\ $}}1
        {--}{{$-\ $}}1
        {\\TT}{{\hspace{1.5em}}}3        
      }

\lstdefinelanguage{pvsNums}[]{pvs}
{
  numbers=left,
  numberstyle=\tiny,
  stepnumber=2,
  numbersep=4pt
}

\lstdefinelanguage{pvsNumsRight}[]{pvs}
{
  numbers=right,
  numberstyle=\tiny,
  stepnumber=2,
  numbersep=4pt
}

\lstnewenvironment{BigPVS}%
  {\lstset{language=BigPVS}}
  {}

\lstnewenvironment{PVSNums}%
  {
  \if@firstcolumn
    \lstset{language=pvs, numbers=left, firstnumber=auto}
  \else
    \lstset{language=pvs, numbers=right, firstnumber=auto}
  \fi
  }
  {}

\lstnewenvironment{PVSNumsRight}%
  {
    \lstset{language=pvs, numbers=right, firstnumber=auto}
  }
  {}

\newcommand{\linefigpvs}[9]{

}

\lstdefinelanguage{pvsproof}{
  basicstyle=\tt \figuresize,
  mathescape=true,
  tabsize=4,
  sensitive=false,
  columns=fullflexible,
  keepspaces=false,
  flexiblecolumns=true,
  basewidth=0.05em,
}

\newcommand{\tuple}[1]{\left\langle#1\right\rangle}

\def\Var{\act{Var}}

\newcommand{\toolhylink}{HyLink\xspace}

\newcommand{\toolspaceex}{SpaceEx\xspace}

\newcommand{\tooldaikon}{{Daikon}\xspace}
\newcommand{\toolframac}{{Frama-C}\xspace}
\newcommand{\toolhynger}{Hynger\xspace}

\def\N{\act{N}}

\def\Var{\act{V}}

\newcommand{\val}[1]{{\mathit{val}(#1)}}
\newcommand{\vars}[1]{{\mathit{vars}(#1)}}

\newcommand{\localvar}[2]{{{#1_{#2}}}}

\def\xi{\localvar{x}{i}}

\def\reach{{\sf Reach}}

\def\Vs{V_S}
\def\Vout{V_{\mathit{out}}}
\def\Vref{V_{\mathit{ref}}}
\def\Vripple{V_{\mathit{err}}}
\def\Vtol{V_{\mathit{tol}}}

\def\specSet{\Sigma}
\def\specSetPhysical{\specSet_P}
\def\specPhysicalSymbol{\sigma_P}
\newcommand{\specPhysical}[1]{\sigma^{#1}_{P}}
\def\specSetCyber{\specSet_C}

\newcommand{\inferredSpecPhysical}[1]{\hat{\varphi}^{#1}_{P}}

\def\HyngerVertices{M}
\def\HyngerEdges{E}
\def\HyngerGraph{G}
\def\HyngerAllVar{\Var}
\newcommand{\HyngerRoot}[1]{\mathit{root}(#1)}
\newcommand{\HyngerChildren}[1]{\mathit{children}(#1)}
\newcommand{\HyngerParent}[1]{\mathit{parent}(#1)}
\newcommand{\HyngerLevel}[1]{\mathit{siblings}(#1)}
\newcommand{\HyngerAncestors}[1]{\mathit{ancestors}(#1)}
\newcommand{\HyngerVar}[1]{\Var({#1})}
\newcommand{\HyngerVarInput}[1]{\Var_I(#1)}
\newcommand{\HyngerVarOutput}[1]{\Var_O(#1)}
\newcommand{\HyngerVarCyber}[1]{\Var_C(#1)}
\newcommand{\HyngerVarPhysical}[1]{\Var_P(#1)}
\def\HyngerVarSP{\Var_{SP}}

\def\HyngerVarGraph{G_{\Var}}
\def\HyngerVarVertices{V_{\Var}}
\def\HyngerVarEdges{E_{\Var}}

\def\HyngerConnectsSymbol{\hookrightarrow}
\newcommand{\HyngerBlockConnects}[2]{#1\HyngerConnectsSymbol#2}
\def\HyngerPathSymbol{\leadsto}
\newcommand{\HyngerVertexPath}[2]{#1\HyngerPathSymbol#2}

\def\Vs{V_S}
\def\Vout{V_{\mathit{out}}}
\def\Vref{V_{\mathit{ref}}}
\def\Vripple{V_{\mathit{rip}}}
\def\Vtol{V_{\mathit{tol}}}
\def\iL{i_L}
\def\Vc{V_C}
\def\iLdot{\dot{i}_L}
\def\Vcdot{\dot{V}_C}

\newcommand{\Invset}{{\mathit{Inv}}}
\newcommand{\Initset}{{\mathit{Init}}}
\newcommand{\Flowset}{{\mathit{Flow}}}

\newcommand{\Locset}{{\mathit{Loc}}}
\newcommand{\Varset}{{\mathit{Var}}}
\newcommand{\Transset}{{\mathit{Trans}}}
\newcommand{\Labelset}{{\mathit{Lab}}}
\newcommand{\Trajectoryset}{{\mathit{Traj}}}

\newcommand{\guard}{g}
\newcommand{\reset}{u}
\newcommand{\VarInput}{{\mathit{I}}}
\newcommand{\VarOutput}{{\mathit{O}}}
\newcommand{\VarCyber}{{\mathit{C}}}
\newcommand{\VarPhysical}{{\mathit{P}}}
\def\AutomatonIO{\tilde{\A}}
\newcommand{\Real}{\mathbb{R}}
\newcommand{\Realn}{\mathbb{R}^n}
\def\LabelActionBuck{\theta}

\def\AFref{\lambda_{\mathit{ref}}}

%% file: abstract.tex
Embedded systems use increasingly complex software and are evolving into cyber-physical systems (CPS) with sophisticated interaction and coupling between physical and computational processes.  Many CPS operate in safety-critical environments and have stringent certification, reliability, and correctness requirements. These systems undergo changes throughout their lifetimes, where either the software or physical hardware is updated in subsequent design iterations. One source of failure in safety-critical CPS is when there are unstated assumptions in either the physical or cyber parts of the system, and new components do not match those assumptions.  In this work, we present an automated method towards identifying unstated assumptions in CPS. Dynamic specifications in the form of candidate invariants of both the software and physical components are identified using dynamic analysis (executing and/or simulating the system implementation or model thereof). A prototype tool called Hynger (for HYbrid iNvariant GEneratoR) was developed that instruments Simulink/Stateflow (SLSF) model diagrams to generate traces in the input format compatible with the Daikon invariant inference tool, which has been extensively applied to software systems.  Hynger, in conjunction with Daikon, is able to detect candidate invariants of several CPS case studies. We use the running example of a DC-to-DC power converter, and demonstrate that Hynger can detect a specification mismatch where a tolerance assumed by the software is violated due to a plant change. Another case study of an automotive control system is also introduced to illustrate the power of Hynger and Daikon in automatically identifying cyber-physical specification mismatches.


%% file: intro.tex
%
\section{Introduction}
\seclabel{intro}
Systems interacting with their physical environments are becoming increasingly dependent upon computers and software, such as in emerging cyber-physical systems (CPS).
For instance, typical modern cars utilize hundreds of microprocessors, many communications buses, and a complex interconnection between sensors, actuators, and processors.
In the design and development process for most engineered systems, the vast majority of resources are devoted to ensuring systems meet their specifications~\cite{beizer1990book}. 
%
However, in spite of significant technical advances for designing verification and validation such as model checking, Software/Hardware-In-The-Loop (SIL/HIL) testing, automatic test case generation for software, and sophisticated simulators, there still remain products recalled across industries for safety concerns due to software problems and system integration between the cyber and physical subcomponents.
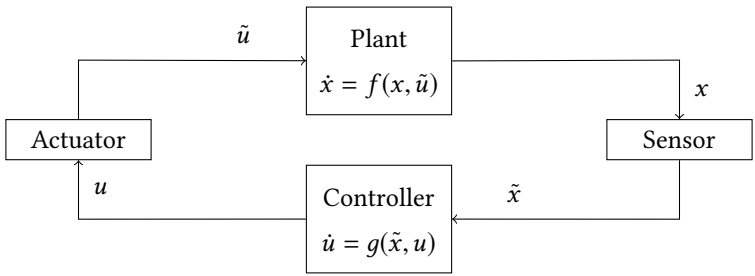
\begin{figure}[t]%
	\centering%
	\begin{tikzpicture}[r/.style={draw,rectangle, minimum height=1.5em, minimum width=5.5em},
		a/.style={draw,->,>=stealth}]
	\node[r] (s1) {\begin{tabular}{@{}c@{}}
			Plant \\
			$\dot{x} = f(x,\tilde{u})$
		\end{tabular}};
	\node[r] (s2) [below = 6.5mm of s1] {\begin{tabular}{@{}c@{}}
			Controller \\
			$\dot{u} = g(\tilde{x}, u)$
		\end{tabular}};
	\node[r] (s3) [below = 0.5mm of s1, xshift= 4cm] {Sensor};
	\node[r] (s4) [below = 0.5mm of s1, xshift= -4cm] {Actuator};
	\node[coordinate] (c1) [above=7.75mm of s4.north] {};
	\node[coordinate] (c2) [below=7.75mm of s4.south] {};
	\node[coordinate] (c3) [above=7.75mm of s3.north] {};
	\node[coordinate] (c4) [below=7.75mm of s3.south] {};
	\draw[->] (s4.north) -- (c1) -- (s1);
	\draw[->] (s2.west) -- (c2) -- (s4);
	\draw[->] (s1.east) -- (c3) -- (s3);
	\draw[->] (s3.south) -- (c4) -- (s2);
	\node (plant_output) [below=2mm of c3,xshift=3mm]{$x$};
	\node (controller_input) [above=1mm of c4,xshift=-22mm]{$\tilde{x}$};
	\node (controller_output) [above=2mm of c2,xshift=3mm]{$u$};
	\node (plant_input) [above=1mm of c1,xshift=22mm]{$\tilde{u}$};
\end{tikzpicture}
	\caption{High-level diagram of a closed-loop control system.}%
	\figlabel{feedback_controller}%
	\vspace{-1em}%
\end{figure}%
The verification community typically focuses on the \emph{developmental verification}, where a model of a system is developed and properties (specifications) are (manually, semi-automatically, or automatically) checked for that system.
However, many product recalls and safety disasters induced by software bugs are not a result of design errors, but are the result of either:
\begin{enumerate*}[label=(\textit{\alph*})]
\item implementation errors, or
\item reuse, upgrade, and maintenance errors.
\end{enumerate*}
Initiatives like a priori Model-Based Design (MBD) are important research efforts and may someday lead to synthesizing provably correct implementations from specifications.
However, most systems being designed today still utilize a development process that has engineers writing the software, and systems are integrated with numerous components in a potentially error-prone process.
%
%
%
For instance, a typical CPS that has been used widely in control systems is a closed-loop feedback controller shown in \figref{feedback_controller}, where a plant describes physical changes of the environment and a controller represents cyber/software computations corresponding to these changes.
The physical evolution of the plant can be sensed and sampled by a sensor, and then fed into the controller.
Based on the measurement of the plant provided by the sensor, the controller provides a corresponding control signal to regulate the physical changes in the plant. This control signal is converted by an actuator before sending it to the plant.
Such a system may contain different possibilities of failure due to the following main reasons:
\begin{enumerate*}[label=(\textit{\alph*})]
\item the controller may make wrong assumptions about the plant, sensor or actuator. For example, changing parameters of the plant, sensor, or actuator without updating the controller may produce potential specification mismatches. This controller-reuse issue can lead to safety failures such as the Honda vehicles recalls or the Ariane 5 flight 501 disaster described in \secref{design_reuse}.
\item The plant may be influenced by uncontrolled factors (disturbances) from the environment,
\item the controller is initially encoded based on wrong information about the physical plant,
\item the sensor and actuator may have conversion errors, and
\item the control conflicts may arise when using a shared sensor and actuator network.
\end{enumerate*}

In this \papertype, we develop a method to address such challenges that arise in the product evolution and upgrade process in CPS.
Our proposed method enables dynamic analysis using well-established software engineering tools for large classes of Simulink/Stateflow (SLSF) models that are frequently used in CPS engineering.
In particular, the method infers candidate invariants of SLSF models.
Invariants are properties of a system that should always hold, while conditional invariants may hold at certain program points, for example, at the beginning or end of a function call (pre/post conditions).
This is important because such models are amenable to formal verification using existing research tools and hybrid system model checkers. Finding invariants can aid this process as they represent potential abstractions with a possibly less complex representation than the set of reachable states.
The SLSF block diagrams may be black box components, white box components, or even system implementations (such as when SLSF is used in SIL/HIL simulation).
In the case when the underlying SLSF models are known, they may be formalized using hybrid automata~\cite{manamcheri2011hscc}.
Candidate invariants inferred with our \toolhynger (for HYbrid iNvariant GEneratoR) software tool in conjunction with Daikon~\cite{ernst2001tse,ernst2007scp} may be formally checked as actual invariants using a hybrid system model checker~\cite{frehse2011cav}.~\figref{method_overview} shows a preliminary overview of our proposed methodology. As a prelude, we just intuitively demonstrate how \toolhynger and Daikon can be used to detect specification mismatches. The proposed framework will be fully presented in \secref{tool}.
%
%
\begin{figure}[t!]%
	\centering%
	\includegraphics[width=0.95\columnwidth]{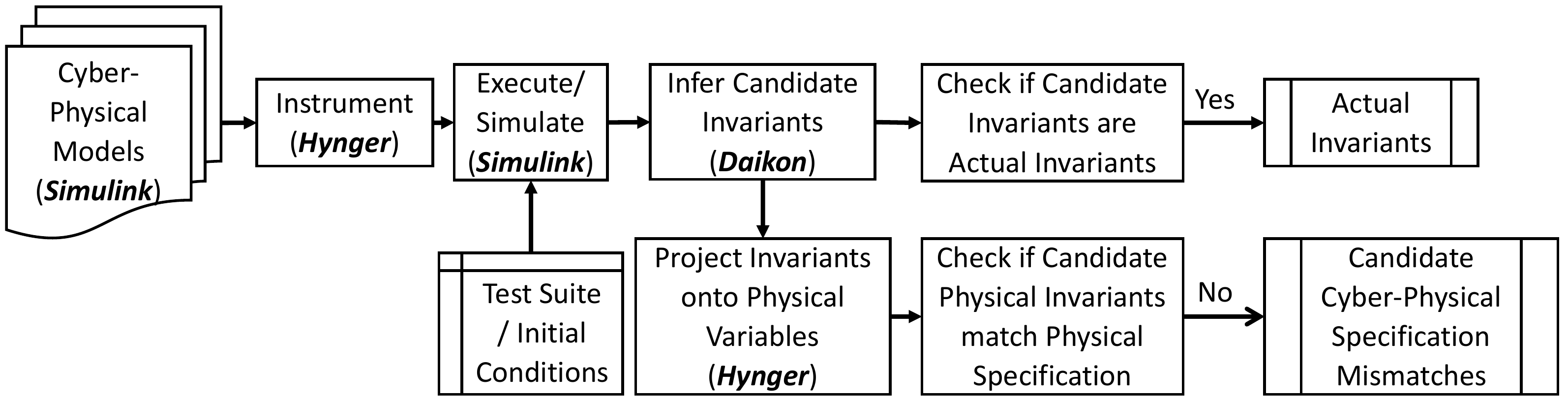}%
	\caption{Preliminary overview of the proposed methodology using \toolhynger and Daikon to infer candidate invariants and detect specification mismatches.}%
	\figlabel{method_overview}%
	\vspace{-1em}%
\end{figure}%





\paragraph*{Contributions}
The primary contributions of this \papertype are:
\begin{enumerate*}[label=\textit{(\alph*)}]
\item the formalization of the cyber-physical specification mismatch problem,
\item a methodology for performing template-based automated invariant inference of white box (known) and black box (unknown) CPS models using dynamic analysis,
\item the \toolhynger software tool, which supports instrumenting large classes of SLSF diagrams for dynamic analysis using tools like Daikon,
\item a methodology for checking if the inferred invariants are actual invariants by using formal models of the underlying SLSF model diagrams and hybrid systems model checkers such as \toolspaceex~\cite{frehse2011cav}, etc.,
\item two proof-of-concept CPS case studies using \toolhynger to automatically identify cyber-physical specification mismatches.
\end{enumerate*}
These results can be used to help bridging the worlds of actual embedded systems software (e.g., detailed SLSF diagrams and generated C code) with hybrid system models.
%
%

Overall, this journal has been substantially extended from our previous work~\cite{johnson2015cyber}. In fact, we added the formal definitions of cyber-physical specification mismatches, cyber-physical input-output automata, and invariant checking problem to identify whether the inferred invariants are actual invariants. Moreover, two proof-of-concept CPS case studies including a buck converter and an abstract fuel control system are presented to show the capability of \toolhynger tool in automatically identifying potential cyber-physical specification mismatches of CPSs. The experimental results illustrate the feasibility of using dynamic invariant inference for analysis of embedded and cyber-physical systems.
Before presenting the details of our approach, we first illustrate the pitfalls of CPS design reuse by citing examples of critical mistakes in existing, certified systems.


%% file: design_reuse.tex
\section{Cyber-Physical Design Reuse and Upgrade}
\seclabel{design_reuse}
In this section, we review cases where CPS design reuse and upgrade have led to failures in existing systems.
This motivates the need for our proposed method and our \toolhynger tool, which can be used to find and formalize unstated assumptions in CPS.
%

\iftoggle{extended_version}
{
When examining recall and safety notices, one theme is that mistakes occurred across multiple fields, across multiple manufacturers, and regardless of the safety-critical (and certified) nature of the system under consideration.

For example,
\begin{enumerate*}[label=\textit{\alph*)}]
\item the Consumer Product Safety Commission (CPSC) has recalled between 2010-2012 fire alarm and control systems from Bosch, Tyco-Grinnel, and Honeywell for failure to sound alarms and/or notify fire departments~\cite{cpsc2010firelite,cpsc2011tyco,cpsc2012bosch},
\item the CPSC has recalled elevators for opening doors without elevator cars present exposing the elevator shaft~\cite{cpsc2012thyssenkrupp},
\item the Food and Drug Administration (FDA) has reported thousands of deaths and hundreds of thousands of injuries caused by computer-related defects in medical devices between 2006 and 2011~\cite{alemzadeh2013sp}, and
\item the National Highway Transportation and Safety Administration (NHTSA) has recalled thousands of Toyota Priuses due to drivetrain software problems causing unexpected stalls~\cite{nhtsa2005toyotaPrius}.
%
%
\end{enumerate*}
These recalls illustrate that we are currently incapable of delivering reliable software systems that interact with the physical world due to the complexity of the software, communications, and physical interactions.
}
A recent example of a design-reuse problem is the National Highway Transportation and Safety Administration (NHTSA) recall of $1.5$ million Honda vehicles (including one of the author's) due to Electronic Control Module (ECM) software problems that could damage the car's transmission, resulting in possible stalls.
The root cause of the safety defect was the result of a physical component (a bearing in the transmission) being upgraded to an improved design between different model-year vehicles without appropriate ECM software updates~\cite{nhtsa2011honda}. This problem was widespread because there was a five year delay before the problem was identified, and it was used across model makes and years (e.g., from $2005-2010$ model year Accords, $2007-2010$ CR-Vs, and $2005-2008$ Elements). This difficulty in root-cause analysis emphasizes the point that such problems are probably underreported, and the reuse of components in CPS can lead to widespread serious problems.

Similar design-reuse problems have famously occurred in aviation---the Ariane 5 flight 501 disaster was a result of reusing Ariane 4's software without appropriate updates for the increased thrust of the new rocket~\cite{lions1996tr,ariane5_flight501}. Here, software developers made an assumption about the physical dynamics of the rocket, but the software was reused from Ariane 4, while Ariane 5 had greater thrust, so this assumption was invalid. Finally, when considering the future of CPS, the Defense Advanced Research Projects Agency's System of Systems Integration Technology and Experimentation (DARPA SoSITE) program~\cite{darpa_sosite} describes modularized military aviation systems which are capable of rapid component swapping and upgrade. Left unaddressed, issues related to unstated assumptions in components are likely to get worse in future CPS, where changes can occur in the software and hardware.

Besides design-reuse problems, software upgrades without being thoroughly tested and validated may result in an epic system failure.
One famous example of this type of problem is the disaster of Mars Climate Orbiter (MCO), developed by NASA's Jet Propulsion Laboratory (JPL).
The root-cause of this disaster was that different parts of the software developer team were using different units of measurements.
In fact, one part of the ground software upgraded by Lockheed Martin Astronautics (LMA) measured the thrusters in English units of pounds (force)-seconds instead of metric units of Newton-seconds as defined in its original Software Interface Specification (SIS) used by JPL~\cite{stephenson1999,leveson2002}.
Therefore, the trajectory of the MCO was erroneously calculated by ground support system computers using the incorrect thruster performance data.
This type of software failure occurred due to the lack of adequate communication between different parts of the software team and the uncovered issues of verification and validation processes~\cite{stephenson1999}.
\input{rw}

%% file: rw.tex
\subsection{Related Work}
\seclabel{rw}
The idea evaluated in this work, that of inferring physical system specifications from embedded software in conjunction with physical system models and evaluating them for mismatches, was inspired by previous work finding program specifications for pure software systems~\cite{nimmer2002issta}.
Cyber-physical specification mismatch is closely related to model inconsistency~\cite{reder2013tse}, architectural mismatch~\cite{garlan1995icse}, and requirements consistency~\cite{whalen2013software}.
There are many benefits of dynamic analysis such as using implementations instead of models~\cite{nimmer2002issta,ernst2001tse,ernst2007scp} to find dynamic program specifications~\cite{nimmer2002issta}, providing documentation over program evolution and checking if specifications change drastically over program evolution, etc.
For one, models are not actually required for analysis, and implementations may be used~\cite{ernst2001tse,ernst2007scp}.
The benefit of executing a system implementation is that there are no mismatches between a model (potentially documentation-based) and implementation, since it is not necessary to have a model at all.
The candidate specification generated may be viewed as a form of input-output abstraction of the actual implementation.
The limitation includes results that are unsound without additional reasoning.

Recently, Medhat and his collaborators introduced a new framework for inferring hybrid automata from black-box implementations of embedded control systems by mining their input/output traces~\cite{Medhat2015}.
In their work, the input/output training traces collected from executing a system are clustered and then translated to event sequences. Under several assumptions, hybrid automata representing the behaviors of the system can be inferred using the input/output correlation.
%
%
Although the work suffers some limitations, their proposed approach is a proof-of-concept of using dynamic analysis to infer the specifications of black-box systems.
This work is highly relevant to our proposed method as there is an analogy between inferring hybrid automata and finding a candidate invariant for a black-box system. In fact, both of them can be considered as doing specification inference using dynamic analysis.

There are also several tools such as DepSys~\cite{munir2014depsys} and EyePhy~\cite{munir2015eyephy} that used both static and dynamic analysis to detect and address the control conflict due to dependencies when using multiple CPS applications.
Particularly, DepSys is a utility sensing and actuation infrastructure for a smart home that can simultaneously operate multiple CPS applications. The main novelty of DepSys is that it provides a comprehensive strategy to specify, detect and automatically address the control conflicts between sensors and actuators used in a home setting.
Similarly, EyePhy is an integrated system that can detect dependencies and then perform a dependency comprehensive analysis across HIL CPS medical applications. A built-in simulator, HumMod, in EyePhy is able to model the complex interactions of the human body using more than 7,800 physiological variables.
HumMod demonstrates the model parameters and the quantitative relationship among them in XML files that makes it easier to update the physiological models without the recompilation of the whole system. EyePhy can be considered as the first tool that performs the dependency analysis across applications' control actions on the human body.
%
%
Additionally, the sensor networks with devices used in smart homes or medical devices can be built using the family of Smart Transducer Interface Standards (IEEE 1451). IEEE 1451 has been developed in order to provide the common communication interfaces for connecting transducers (sensors or actuators) to their instrumentation systems or control networks~\cite{lee2000ieee}.
The Transducer Electronic Data Sheets (TEDS) embedded in smart transducers are memory devices, which store the manufacture-related information of the transducer such as manufacture ID, measurement ranges, serial number, etc. Thus, TEDS allows transducers to be self-identified and self-descriptive to the device networks. It also provides a standardized mechanism to facilitate the plug and play of transducers with different control networks.
Hence, IEEE 1451 enables the access of transducer data through a common set of interfaces, allowing users to select transducers and networks for their applications. This advantage is crucial in facilitating the device and data interoperability, detecting and resolving conflicts due to dependencies when concurrently using multiple transducers in the device networks.

Finding specifications is a maturing field within software engineering~\cite{nimmer2002issta,ernst2001tse,ernst2007scp,boshernitsan2006issta,csallner2008icse}.
Daikon, which is used by \toolhynger, processes program traces to generate invariants~\cite{ernst2001tse,ernst2007scp}.
For several languages (C, C++, etc.), this process is performed without access to the source code by instrumenting the compiled program using Valgrind~\cite{nethercote2007pldi}.
This makes it difficult to use on non-x86/x86-64 platforms (although Valgrind is gaining access to other architectures), which is a serious limitation, as most embedded platforms utilize other architectures (e.g., ARM, AVR, PIC, 8051, MSP430, etc.).
Due in part to these limitations, \toolhynger instruments architecture-independent SLSF diagrams directly.
%
In the long run, the \toolhynger tool is envisioned to take an arbitrary SLSF model, instrument it, then analyze the resulting traces with dynamic analysis to identify broad classes of cyber-physical specification mismatches.

The most closely related work using Daikon is to find candidate invariants of hybrid models of biological system~\cite{bernardini2007wmcs}, and this also illustrates a proof-of-concept of using Daikon as a trace analyzer for non-purely software systems.
Daikon can generate invariants of many forms for variables and data structures, such as: ranges ($a \leq x \leq b$), linear ($y = ax + b$), variable ordering ($x \leq y$), sortedness of lists, etc.
Daikon works by instrumenting source code and/or compiled binaries with changes that allow for looking at variable values, then Daikon essentially checks if variables satisfy some template invariants.
For instance, if an integer variable $x$ is observed to always be smaller than some number, say $50$, Daikon may generate a candidate invariant of $x \leq 50$. 
Based on many advantages of using Daikon as a trace analyzer~\cite{ernst2001tse,ernst2007scp}, we prefer to use \toolhynger with Daikon to infer candidate invariants in our proposed framework. However, we note that \toolhynger can generate a trace file in many input formats that are compatible with other invariant-inference tools using dynamic analysis not just Daikon.
Other research tools like DySy~\cite{csallner2008icse} and commercial tools like Agitagor~\cite{boshernitsan2006issta} can be used for generating candidate invariants for other languages.

%% file: model.tex
\section{Cyber-Physical System Models}
\seclabel{model}
The approach presented in this \papertype applies to the systems with formal models, informal models, and unknown models/implementations.
The primary assumption is that interfaces to the models or systems are available as SLSF blocks.
%
%
There are two main categories of blocks appearing in an SLSF diagram that are supported by our method, white box and black box systems.
The white box systems may contain:
\begin{enumerate*}[label=(\textit{\alph*})]
\item known, informal models,
\item known, informal implementations, or
\item known, formal models (e.g., hybrid automata, or more precisely, classes of SLSF diagrams that may be converted to hybrid automata~\cite{manamcheri2011hscc}).
\end{enumerate*}
The black box systems may be completely unknown, and may contain:
\begin{enumerate*}[label=(\textit{\alph*})]
\item unknown implementations (e.g., compiled executable binaries with no source available, such as commercial off-the-shelf [COTS] components and other third-party systems),
\item unknown models, and
\item actual cyber-physical systems (e.g., embedded controllers, networked computers, and physical plants, all that may show up in HIL/SIL simulations interfaced with SLSF).
\end{enumerate*}
%

%
\todo{assume guarantee reasoning with hierarchical hybrid automata: idea to use internal hierarchy abstracted as just assume and guarantee behaviors, maybe detected using the invariant inference methodology}
\todo{cyber-physical automata}
\todo{physical variables: specific set of variables in software corresponding to physical state; the transitive closure of any variable that uses a value from a physical variable}
\todo{Really need to define some model to be able to clearly talk about. I propose either HIOA quickly, or maybe a simpler formalism, where we don't even define semantics, but instead just define syntax / structure.}
Next, we define a structure of CPS models used in SLSF.
We will not define formal semantics of this structure or SLSF diagrams in this paper.
However, in the case where an SLSF diagram is a white box and formal semantics may be defined, a formal framework like hybrid input/output automata (HIOA)~\cite{lynch2003ic} may be used to specify the semantics, such as done in the \toolhylink tool~\cite{manamcheri2011hscc}.
Additionally, if an SLSF diagram is a white box and linear, we may also be able to use SL2SX Translator for transforming it into a corresponding formal model~\cite{minopoli2016sl2sx}. Interested readers can find some graphical examples of the translation in~\cite{manamcheri2011hscc, minopoli2016sl2sx}.
Other formalisms like actors and hierarchical state machines are commonly used for formal modeling of other diagrammatic frameworks similar to SLSF~\cite{alur2003pi,zhou2012deds,tripakis2013mscs,bensalem2014ssm}.
Given a formal model $\A$ and candidate specification $\Sigma$ (e.g., found using \toolhynger), we can check if $\A$ meets the specification, i.e., $\A \models \specSet$ by using a hybrid system model checker like SpaceEx~\cite{frehse2011cav}.
In some instances, we know when an SLSF diagram corresponds precisely to a hybrid automaton model~\cite{manamcheri2011hscc}, and in these cases, we can check if candidate invariants found with \toolhynger are actual invariants.

First, we define the hierarchy represented by SLSF diagrams. 
\begin{definition}[SLSF diagram]
An SLSF diagram is a tuple $\A \deq \langle \HyngerVertices, \HyngerEdges, \Var \rangle$, where:
\begin{itemize}
\item $\HyngerVertices$ is a set of blocks (vertices) that represent block diagrams (and sub-blocks/models),
\item $\HyngerEdges \subseteq \HyngerVertices \times \HyngerVertices$ is a set of edges between blocks representing a parent-child hierarchy, and
\item $\Var$ is a set of variables, written as $\HyngerAllVar \deq \bigcup_{v \in \HyngerVertices} \HyngerVar{v}$, where $\HyngerVar{v}$ is a set of variables for each block $v \in \HyngerVertices$.
\end{itemize}
\deflabel{slsf_diagram}
\end{definition}
%
According to \defref{slsf_diagram}, the graph $\HyngerGraph \deq (\HyngerVertices,\HyngerEdges)$ defined by the vertices (blocks) $\HyngerVertices$ and edges $\HyngerEdges$ is a rooted tree, where $\HyngerVertices$ are block diagrams and $\HyngerEdges$ represents a parent-child hierarchical relationship (e.g., sub-modules and sub-blocks).
Here, the root (i.e., top-level) block diagram of the model is the unique root of the tree, which we denote as $\HyngerRoot{\HyngerVertices}$.
For a block $v \in \HyngerVertices$, the \emph{children} of $v$ are denoted as $\HyngerChildren{v}$ and defined as the set of blocks $\{ w \in \HyngerVertices \ | \ w \in \HyngerEdges(v) \}$.
For a block $v \in \HyngerVertices$, the \emph{parent} of $v$ is denoted as $\HyngerParent{v}$ and is defined as the singleton set $\{ w \in \HyngerVertices \ | \ v \in \HyngerChildren{w} \}$.
Clearly, $\HyngerParent{\HyngerRoot{\HyngerVertices}} = \emptyset$.
For a block $v \in \HyngerVertices$, the \emph{ancestors} of $v$ are denoted as $\HyngerAncestors{v}$ and defined inductively as the set of blocks $\{ w \in \HyngerVertices \ | \ v \ \cup \ w \in \HyngerChildren{v} \ \cup \ \HyngerChildren{w}  \}$ (or equivalently, as the transitive closure of $\HyngerChildren{v}$).

For a block $v \in \HyngerVertices$, the set of variables of $v$ is $\HyngerVar{v}$ and is partitioned into sets of input and output variables, written respectively as $\HyngerVarInput{v}$ and $\HyngerVarOutput{v}$, and we have $\HyngerVar{v} = \HyngerVarInput{v} \cup \HyngerVarOutput{v}$.
A \emph{variable} $x \in \HyngerVar{v}$ is a name for referring to some state of $\A$, and is associated with a data type denoted $\type{x}$.
Typical data types are reals, floating points, arrays, lists, etc.
The valuation of a variable $x \in \HyngerVar{v}$ is the set of all values it may take and is denoted $\val{x}$.
The state-space of $\A$ is the set of valuations of all the variables $\HyngerAllVar$.
An element $s$ of the state-space is called a state, and a trace is a sequence of states.
The SLSF diagram may also have internal (local) variables, although they are not externally visible, so we do not include them, as only input/output interfaces are visible for external observation and instrumentation.

Next, we define CPS models that appear in SLSF diagrams.
\begin{definition}[CPS model]
A CPS model is an SLSF diagram with a set of $n$ typed variables, $\Var = \{x_1, x_2, \ldots, x_n\}$, which is classified into two subsets as follows.
\begin{itemize}
\item $\Var_P = \{\alpha_1, \alpha_2, \ldots,\alpha_{n_p}\}$ is a set of $n_p \leq n$ physical variables such that their values are continuously updated, and
%
\item $\Var_C = \{\beta_1, \beta_2, \ldots, \beta_{n_c}\}$ is a set of $n_c$ cyber variables that are discretely updated, where $n = n_p + n_c$.
\end{itemize}
\end{definition}
Here, the set of variables for each block of a CPS model is also partitioned into sets of physical and cyber variables, $\HyngerVar{v} = \HyngerVarPhysical{v} \cup \HyngerVarCyber{v}$.
In practice, this may be accomplished with subtyping using, for example, an overloaded type for floats or fixed points used for approximations of real variables (e.g., in C, \texttt{typedef double physical;} \texttt{typedef physical temperature;}).
The dynamic changes of the variables of the CPS model may be described using different SLSF blocks such as S-Function block, look-up table, etc.
%
%
In case the CPS model is a white-box and simple enough, we may translate it to a formal framework like HIOA (e.g using Hylink).
In fact, we can specify a set of real-valued variables and their dynamic changes for the converted formal model based on capturing the output variables from unit delay, integrator, state-space blocks in the corresponding SLSF diagram~\cite{alur2008symbolic}.
Moreover, we note that the input and output variables are disjoint, and the cyber and physical variables are disjoint, although these are not all mutually disjoint. Hence, we further classify the set of variables $\HyngerVar{v}$ into different types as follows.
\begin{definition}[Variable Classification]
For a block $v \in \HyngerVertices$, a variable $x \in \HyngerVar{v}$ is considered as:
\begin{itemize}
\item an \emph{input cyber variable} if $x \in \HyngerVarCyber{v}$ and $x \in \HyngerVarInput{v}$,
\item an \emph{output cyber variable} if $x \in \HyngerVarCyber{v}$ and $x \in \HyngerVarOutput{v}$,
\item an \emph{input physical variable} if $x \in \HyngerVarPhysical{v}$ and $x \in \HyngerVarInput{v}$, or
\item an \emph{output physical variable} if $x \in \HyngerVarPhysical{v}$ and $x \in \HyngerVarOutput{v}$.
\end{itemize}
\deflabel{variable type}
\end{definition}
%
We extend these notations in \defref{variable type} naturally to sets of variables if \emph{all} variables in a set of variables fall into these classes, and will reference them as such.
An arbitrary set of variables may not be mutually disjoint from each of the input, output, cyber, and physical variables.
Thus, for a set of variables $X \subseteq \HyngerAllVar$, we say:
\begin{enumerate*}[label=(\textit{\alph*})]
\item $X$ is \emph{cyber-physical} if there exist both cyber and physical variables in $X$,
\item $X$ is \emph{input-output} if there exist both input and output variables in $X$, and
\item $X$ is \emph{cyber input-output}, \emph{physical input-output}, \emph{cyber-physical input}, or \emph{cyber-physical output} for the other natural permutations.
\end{enumerate*}

Next, using these variable classes, we define classes of SLSF blocks appearing in SLSF diagrams.
For a block $v \in \HyngerVertices$, we say:
\begin{enumerate*}[label=(\textit{\alph*})]
\item $v$ is a \emph{cyber-physical} block if there exist both cyber and physical variables in $\HyngerVar{v}$,
\item $v$ is a \emph{cyber} block if there exist \emph{only} cyber variables in $\HyngerVar{v}$, and
\item $v$ is a \emph{physical} block if there exist \emph{only} physical variables in $\HyngerVar{v}$.
\end{enumerate*}

\paragraph*{Cyber-Physical Variable Interactions}
Next, we will formalize a notion of influence between cyber and physical models and their variables.
For example, consider a typical closed-loop plant-controller architecture, where outputs of a plant are sensed, used as inputs to a controller, and outputs of the controller are converted by actuators as inputs to the plant (and potentially disturbances affect everything).
Generally, we would say the plant is a physical model, the controller is a cyber model, and the sensors and actuators are cyber-physical models.
However, it is clear that the physical variables of the plant affect the cyber variables of the controller, and vice-versa, albeit not directly, but through the transitive closure of input-output connections over all blocks in the SLSF diagram.
%
%
We note that this is related to the notion of tainted variables in program analysis that is popular in security~\cite{schwartz2010sp}.
To formalize this notion, we specify interconnections between input and output variables between blocks $v \in \HyngerVertices$ at the same hierarchical level in the diagram.

Input-output connections may only exist between models with the same parent (i.e., those in the same hierarchical structure).
%
%
For a block $v \in \HyngerVertices$, we denote all blocks with the same parent as $\HyngerLevel{v}$, which is defined as the set $\{ w \in \HyngerVertices \ | \ \HyngerParent{w} = \HyngerParent{v} \}$.
Output variables of a block $v \in \HyngerVertices$ may be connected to input variables of a block $w \in \HyngerVertices$.
\todo{parameterized next definition on block v? probably so.}
Let $\HyngerVarGraph \deq (\HyngerVarVertices, \HyngerVarEdges)$ be a directed graph where the vertices $\HyngerVarVertices$ are variables of blocks $v \in \HyngerVertices$ and the edges specify the interconnection between output variables to input variables for some model $w \in \HyngerLevel{v}$, and we have $\HyngerVarEdges \subseteq \HyngerVar{v} \times \HyngerVar{w}$.
In general, for a fixed block $v \in \HyngerVertices$ and variable $x \in \HyngerVar{v}$, this interconnection relation is a tree, rooted at the output variable $x$ and connected to possibly many input variables of other blocks $w \in \HyngerVertices$ for $w \neq v$.
For two blocks $v, w \in \HyngerVertices$, we say $v$ \emph{connects to} $w$ if there exists an output variable $y \in \HyngerVarOutput{v}$ and an input variable $u \in \HyngerVarInput{w}$ with $\HyngerVarEdges(u) = y$, denoted $\HyngerBlockConnects{v}{w}$.
For two blocks $v, w \in \HyngerVertices$, we say $v$ \emph{has a path to} $w$ if $w$ is in the transitive closure of blocks that $v$ connects to (i.e., $v \HyngerConnectsSymbol^* w$), denoted $\HyngerVertexPath{v}{w}$.
We note that the $\HyngerPathSymbol$ relation may have cycles, and such cases arise in feedback control loops.
For a block $v \in \HyngerVertices$, for an input variable $u \in \HyngerVarInput{v}$ and output variable $y \in \HyngerVarOutput{v}$, we say $u$ \emph{directly influences} $y$ if the value of $y$ changes as a function of $u$.\footnote{Internally the blocks may be very sophisticated, could represent complex physical systems, could be Turing complete, etc., so we use this abstract notion.}
Finally, for two blocks $v, w \in \HyngerVertices$ such that $\HyngerVertexPath{v}{w}$, for an output variable $y \in \HyngerVarOutput{v}$ and an input variable $u \in \HyngerVarInput{w}$, we say $y$ \emph{influences} $u$ if there exists a sequence of directly influenced variables between $y$ and $u$.
%
%
Thus, we can see that a cyber variable in one model may influence a physical variable in another model (or even its own model if there is a cycle), and vice-versa.
The \emph{software physical variables} are all cyber variables that are influenced by physical variables, and are denoted $\HyngerVarSP$.
Typical examples of software physical variables include those used for sensed and sampled measurements, variables used in feedback control calculations, etc.

\begin{example}
Here, we describe a CPS case study used throughout the remainder of the \papertype for illustrating concepts.
The case study is a DC-to-DC power converter (like buck, boost, and buck-boost converters)~\cite{nguyen2014arch}, all of which have similar modeling, but we focus particularly on a buck converter.
The buck converter has two real-valued state variables modeling the inductor current $i_L$ and the capacitor voltage $V_C$.
These state variables are written in vector form as: $x = \left[ i_L; V_C \right]$.
The dynamics of the continuous variables in each mode $m \in \{\mathit{Open}, \mathit{Close}, \mathit{DCM}\}$ are specified as linear (affine) differential equations: $\dot{x} = A_m x + B_m u$, where $u = \Vs$ is a source voltage.
The $A_m$ matrices consist of $L > 0$, $R > 0$, $C > 0$ real-valued constants, respectively representing inductance (in Henries), resistance (in Ohms), and capacitance (in Farads).
A buck converter takes an input voltage of say $5$V and ``bucks'' or drops the voltage to some lower DC voltage, say $2.5$V.
These circuits are used in many electronic devices (e.g., personal computers, cellphones, embedded systems, aircraft, satellites, cars).
These circuits are also used as modular components in a variety of novel power electronics architectures, such as AC/DC microgrids and distributed DC-to-AC multilevel inverters~\cite{nguyen2014tec}.
The general architecture of the buck converter that we focus on consists of a plant (physical system) model and a controller (cyber model/software), along with models of actuators and sensors interfacing the plant and controller.
A controller for the buck converter may be constructed as a hysteresis controller, which changes the mode of the buck converter plant based on the measured output voltage~\cite{hossain2013peci}.
In fact, the converter is meant to transform a given source voltage $\Vs$ to create an output voltage $\Vout$ approximately equal to a desired reference voltage (or set-point) $\Vref$.
To accomplish this, the switch controlling whether $\Vs$ is connected to the output or not is toggled depending on whether $\Vout > \Vref$ or $\Vout < \Vref$.
In practice, to avoid switching too often, a hysteresis band is used and switches occur when $\Vout > \Vref + \Vtol$ or $\Vout < \Vref - \Vtol$.
The choice of $\Vtol$, along with the system dynamics, will determine the voltage ripple $\Vripple$ about the set-point $\Vref$.
Typical specifications require the voltage ripple to be small, so that the output voltage $\Vout$ is approximately $\Vref$, that is, $\Vripple$ is chosen so that for $\Vout = \Vref \pm \Vripple$, we have $\Vout \approx \Vref$.
%
%
The sensor model performs quantization and sampling, as would occur in typical analog to digital conversion (ADC) used to digitize analog signal measurements.
The actuator models likewise perform the inverse process of digital to analog conversion (DAC) to convert the digital (cyber) signals to analog signals.
%

Generally, we can model the plant as a physical block, the hysteresis controller as a cyber block, and the sensors and actuators as cyber-physical blocks in SLSF.
The plant voltage is an output physical variable that affects the output cyber variable---a switching mode signal that enables the transition between each mode in the plant---of the controller, and vice-versa.
This interaction between the plant and the controller is accomplished through the transitive closure of input-output connections with the sensor and the actuator in the SLSF model.
We will formalize specifications and mismatches of the buck converter in~\secref{spec_mis}.
As a prelude, we highlight that \toolhynger finds its candidate invariant (that can be shown to be an actual invariant when modeled as a hybrid automaton~\cite{johnson2012peci,hossain2013peci,nguyen2014arch}).
%
%
%
%
%
\end{example}

\def\BuckStateVector{\left[ \begin{array}{c} \iL \\ \Vc \end{array} \right]}
\def\BuckStateVectorDot{\left[ \begin{array}{c} \iLdot \\ \Vcdot \end{array} \right]}

%
%

\vspace{-1em}
\subsection{Cyber-Physical Input-Output Automata}
\seclabel{cpioha}
To further investigate cyber-physical specification mismatches of CPS models, we consider ones that may be formally represented as cyber-physical input-output automata.
\vspace{-1em}
\begin{definition}
\deflabel{Automaton}
A cyber-physical input-output automaton (CPIOA) $\AutomatonIO$ is a tuple, $\AutomatonIO$ $\deq$ $\langle$$\Locset$, $\Varset$, $\Flowset$, $\Invset$, $\Trajectoryset$, $\Labelset$, $\Transset$, $\Initset$$\rangle$, consisting of the following components:
\begin{itemize}
\item $\Locset$: a finite set of discrete locations.
\item $\Varset$: a finite set of $n$ continuous, real-valued variables, where $\forall x \in \Varset$, $\val{x} \in \Real$ and $\val{x}$ is a valuation---a function mapping $x$ to a point in its type---here, $\Real$; and $\Q \deq \Locset \times \Realn$ is the state space. $\Varset$ is the disjoint of a set of input variables $\VarInput$ and a set of output variables $\VarOutput$. Furthermore, $\VarCyber$ and $\VarPhysical$ classify $\Varset$ into sets of cyber and physical variables, respectively.
\item $\Invset$: a finite set of invariants for each discrete location, $\forall \ell \in \Locset$, $\Invset(\ell) \subseteq \Realn$.
\item $\Flowset$: a finite set of derivatives for each continuous variable $x \in \Varset$, and $\Flowset(\ell, x) \subseteq \Realn$ describes the continuous dynamics of each location $\ell \in \Locset$. if $x$ is a physical variable, $\Flowset(\ell, x)$ is a non-zero Lipschitz continuous differential equation over time. Otherwise, if $x$ is a cyber variable, $\Flowset(\ell, x) = 0$.
\item $\Trajectoryset$: a finite set of continuous trajectory models the valuations of variables over an interval of real time $[0, T]$. Let $\Delta_0$, $\Delta_t$ and $\Delta_T$ be the valuations of variable $x$ at time points $0$, $t$, and $T$ respectively, $\forall t \in [0, T]$, $\forall x \in \Varset$, $\exists \ell \in \Locset$, a trajectory $\tau \in \Trajectoryset$ is a mapping function $\tau: [0, T] \rightarrow \val{\Varset}$ such that:
\begin{itemize}
\item $\Delta_t = \Delta_0 + \int_{\delta = 0}^{t} \Flowset(\ell, x) d\delta$, and 
\item $\Delta_0 \models \Invset(\ell)$, $\Delta_t \models \Invset(\ell)$, and $\Delta_T \models \Invset(\ell)$.
\end{itemize}
%
%
%

%
%
%
%
\item $\Labelset$: a finite set of synchronization labels.
\item $\Transset$: a finite set of transitions between locations; each transition is a tuple $\gamma \deq \tuple{\ell, \ell', \guard, \reset}$, which can be taken from source location $\ell$ to destination location $\ell'$ when a guard condition $\guard$ is satisfied, and the post-state is updated by an update map $\reset$.
\item $\Initset$ is an initial condition, which consists of a set of locations in $\Locset$ and a formula over $\Varset$, so that $\Initset \subseteq \Q$.
\end{itemize}
\end {definition}
Next, we define the semantics of a CPIOA $\AutomatonIO$ in term of executions. An execution of $\AutomatonIO$ is a sequence of states, written as $\rho \deq s_0 \rightarrow s_1 \rightarrow s_2 \rightarrow \ldots$, where $s_0 \in \Initset$, and $s_i \rightarrow s_{i+1}$ is the update from the current-state $s_i$ to the post-state $s_{i+1}$, that is specified by the transition relations of the CPIOA $\AutomatonIO$ including:
\begin{inparaenum}[(a)]
\item a discrete transition that demonstrates the instantaneous state update, or
\item a continuous trajectory that represents the state update over a real time interval.
\end{inparaenum}
We say a state $s_k$ is reachable from an initial state $s_0$ if there exists an execution $\rho \deq s_0 \rightarrow s_1 \rightarrow \ldots \rightarrow s_k$.
\paragraph*{Invariant Property} An \emph{invariant property} $\varphi$ of a CPIOA $\AutomatonIO$ is a formula over $\Varset$ and $\Locset$ that is always true for every reachable state of $\AutomatonIO$. Formally, we say $\AutomatonIO \models \varphi$ iff $\forall s \in Reach(\AutomatonIO)$, $s \models \varphi$, where $Reach(\AutomatonIO)$ denotes the set of reachable states of $\AutomatonIO$.
\paragraph*{Parallel Composition} Consider two CPIOAs $\AutomatonIO_1$ $\deq$ $\langle$$\Locset_1$, $\Varset_1$, $\Invset_1$, $\Flowset_1$, $\Trajectoryset_1$, $\Labelset_1$, $\Transset_1$, $\Initset_1$$\rangle$, and $\AutomatonIO_2$ $\deq$ $\langle$$\Locset_2$, $\Varset_2$, $\Invset_2$, $\Flowset_2$, $\Trajectoryset_2$,  $\Labelset_2$, $\Transset_2$, $\Initset_2$$\rangle$, we consider that $\AutomatonIO_1$ and $\AutomatonIO_2$ is \emph{compatible} if
\begin{enumerate*}[label=(\textit{\alph*})]
\item $\VarInput_1 \subseteq \VarOutput_2$,
\item $\VarInput_2 \subseteq \VarOutput_1$, and
\item $\VarOutput_1 \cap \VarOutput_2 = \emptyset$.
\end{enumerate*}
The parallel composition operation combines two compatible CPIOAs into a single CPIOA that represents the synchronous interaction between these two CPIOA when running simultaneously. 
\begin{definition}[Parallel Composition]
Given two compatible CPIOAs $\AutomatonIO_1$ and $\AutomatonIO_2$, the parallel composition of $\AutomatonIO_1$ and $\AutomatonIO_2$ is a CPIOA $\AutomatonIO$ , written as $\AutomatonIO$ $\deq$ $\AutomatonIO_1 \| \AutomatonIO_2$, where:
\begin{itemize}
\item $\Locset = \Locset_1 \times \Locset_2$,
\item $\Varset = \Varset_1 \cup \Varset_2$,
\item $\Q = \Q_1 \times \Q_2$,
\item $\VarOutput = \VarOutput_1 \cup \VarOutput_2$,
\item $\VarInput = (\VarInput_1 \cup \VarInput_2) \setminus \VarOutput$,
\item $\forall \ell_1, \ell_2 \in \Locset$, $\Invset(\ell_1, \ell_2) = \Invset_1(\ell_1) \wedge \Invset_2(\ell_2)$
\item $\forall \ell_1, \ell_2 \in \Locset$, $\forall x \in \Varset$, $((\ell_1, \ell_2), \val{x} \in \Initset)$ iff $(\ell_1, \val{x}) \in \Initset_1 \wedge (\ell_2, \val{x}) \in \Initset_2$,
\item $\Labelset = \Labelset_1 \cup \Labelset_2$,
\item $\forall i \in \{1,2\}$, there is a trajectory $\tau \in \Trajectoryset$ iff $\tau \downarrow (\Locset_i \cup \Varset_i) \in \Trajectoryset_i$, where $\tau \downarrow (\Locset_i \cup \Varset_i)$ denotes the projection of $\tau$ onto the sets of variables and locations of component $i$.
\item Given $\gamma_1 \in \Transset_1$, $\gamma_1 \deq \tuple{\ell_1, \ell'_1, \guard_1, \reset_1}$ and $\gamma_2 \in \Transset_2$, $\gamma_2 \deq \tuple{\ell_2, \ell'_2, \guard_2, \reset_2}$, there exists a transition $\gamma \in \Transset$, $\gamma \deq \tuple{\ell, \ell', \guard, \reset}$ iff:
\begin{itemize}
\item $\ell = (\ell_1, \ell_2)$, $\ell' = (\ell'_1, \ell_2)$, $\guard = \guard_1$, and $\reset = \reset_1$, or
\item $\ell = (\ell_1, \ell_2)$, $\ell' = (\ell_1, \ell'_2)$, $\guard = \guard_2$, and $\reset = \reset_2$, or
\item $\ell = (\ell_1, \ell_2)$, $\ell' = (\ell'_1, \ell'_2)$, $\guard = \guard_1 \wedge \guard_2$, and $\reset = \reset_1 \cup \reset_2$.
\end {itemize}
\end{itemize}
\deflabel{parallel composition}
\end{definition}

\paragraph*{Closed-loop CPIOA} One type of CPS model that we focus on in this paper is a closed-loop model, e.g., the closed-loop buck converter. 
Such a model can be formally represented as a closed-loop CPIOA, which is a parallel composition of a plant and controller CPIOA. The plant CPIOA has continuous dynamics modeled by ordinary differential equations, and the controller CPIOA can be purely discrete.
For instance, the hybrid automaton of the closed-loop buck converter (without sensor and actuator) shown in \figref{automaton_buckboost} can be considered as one closed-loop CPIOA, where $\LabelActionBuck$ is a synchronization label and $mode$ is a discrete control signal. The capacitor voltage variable $V_C$ is not only an output physical variable for the plant CPIOA, but is also an input cyber variable of the controller CPIOA.
In this case, we can check whether the candidate invariants of the closed-loop buck converter found with Hynger and Daikon are actual invariants by investigating its formal model (e.g., a closed-loop CPIOA shown in \figref{automaton_buckboost}) using some hybrid systems model checkers such as \toolspaceex~\cite{frehse2011cav}.
%
%
%
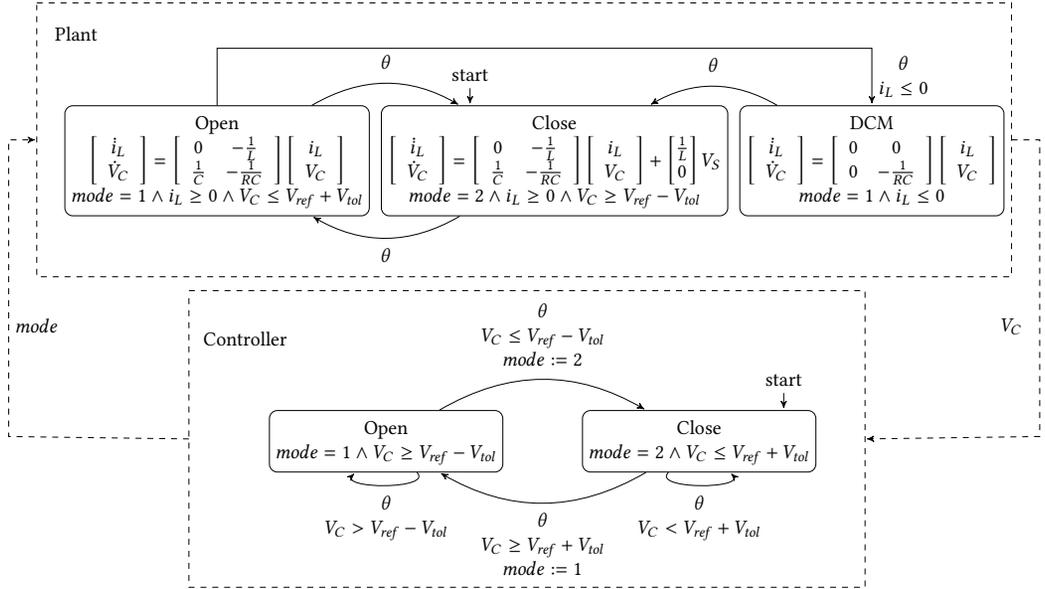
\begin{figure}[t!]%
	\centering%
	\scalebox{0.75}{
	\begin{tikzpicture}[>=stealth',shorten >=1pt,auto,node distance=9cm,font=\normalsize]
		\tikzstyle{every state}=[font=\normalsize,rectangle,rounded corners]
		\node	[state]		(open) 	{\makecell[c]{$\text{Open}$\\$\BuckStateVectorDot = \left[ \begin{array}{cc} 0 & -\frac{1}{L} \\ \frac{1}{C} & -\frac{1}{R C} \end{array} \right] \BuckStateVector$\\$mode = 1 \wedge \iL \geq 0 \wedge \Vc \leq \Vref + \Vtol$}};
		\node[state] (closed)      	[right=2mm of open]		{\makecell[c]{$\text{Close}$\\$\BuckStateVectorDot = \left[ \begin{array}{cc} 0 & -\frac{1}{L} \\ \frac{1}{C} & -\frac{1}{R C} \end{array} \right] \BuckStateVector + \left[ \begin{matrix} \frac{1}{L}\\ 0 \end{matrix} \right] \Vs$\\$mode = 2 \wedge \iL \geq 0 \wedge \Vc \geq \Vref - \Vtol$}};
		\node	[state]		(dcm) [right=2mm of closed]	{\makecell[c]{$\text{DCM}$\\$\BuckStateVectorDot = \left[ \begin{array}{cc} 0 & 0 \\ 0 & -\frac{1}{R C} \end{array} \right] \BuckStateVector$\\$mode = 1 \wedge \iL \leq 0$}};
		\path[->]			(closed)	edge[bend left] node[below] {\makecell[c]{$\LabelActionBuck$}} (open);
		\path[->]			(open)	edge[bend left] node[above] {\makecell[c]{$\LabelActionBuck$}} (closed);
		\node[coordinate] (c1) [above=10mm of open.north] {};
		\node[coordinate] (c2) [above=10mm of dcm.north] {};
		\draw[->]			(open.north) -- (c1) -- (c2) to node[right] {\makecell[c]{$\LabelActionBuck$\\$\iL \leq 0$}} (dcm);
		\path[->]			(dcm)	edge[bend right] node[above]{\makecell[c]{$\LabelActionBuck$}} (closed);
		\node (init) [above=3mm of closed,xshift=-15mm] {start};
		\draw[->] (init) -- (open.north -| init);
		\node (plant) [above=10mm of open,xshift=-25mm] {Plant};
		\node [coordinate] (p1) [above=3mm of plant,xshift=-7mm]{};
		\node [coordinate] (p2) [above=3mm of plant,xshift=166mm]{};
		\node [coordinate] (p3) [below=48mm of p2]{};
		\node [coordinate] (p4) [below=48mm of p1]{};
		\node [coordinate] (pmid23) [below=24mm of p2]{};
		\node [coordinate] (pmid23r) [right=5mm of pmid23]{};
		\node [coordinate] (pmid14) [below=24mm of p1]{};
		\node [coordinate] (pmid14l) [left=5mm of pmid14]{};
		\draw [dashed] (p1)-- (p2) -- (p3) -- (p4)-- (p1){};

		\node	[state]		(controller_open)[below=34mm of open.south, xshift = 30mm]{\makecell[c]{$\text{Open}$\\$mode = 1 \wedge \Vc \geq \Vref - \Vtol$}};
		\node	[state]		(controller_closed) [right=14mm of controller_open]	{\makecell[c]{$\text{Close}$\\$mode = 2 \wedge \Vc \leq \Vref + \Vtol$}};
		\path[->]			(controller_open)	edge[bend left] node[above] {\makecell[c]{$\LabelActionBuck$\\$\Vc \leq \Vref - \Vtol$\\$mode := 2$}} (controller_closed);
		\path[->]			(controller_closed)	edge[bend left] node[below] {\makecell[c]{$\LabelActionBuck$\\$\Vc \geq \Vref + \Vtol$\\$mode := 1$}} (controller_open);
		\path[->]			(controller_open)	edge[loop below, in=225, out=315] node{\makecell[c]{$\LabelActionBuck$\\$\Vc > \Vref - \Vtol$}} (controller_open);
		\path[->]			(controller_closed)	edge[loop below, out=225, in=315] node{\makecell[c]{$\LabelActionBuck$\\$\Vc < \Vref + \Vtol$}} (controller_closed);
		\node (init) [above=3mm of controller_closed,xshift= 15mm] {start};
		\draw[->] (init) -- (controller_closed.north -| init);
		\node (controller) [above=10mm of controller_open, xshift=-25mm] {Controller};
		\node [coordinate] (co1) [above=6mm of controller,xshift=-10mm]{};
		\node [coordinate] (co2) [above=6mm of controller,xshift=110mm]{};
		\node [coordinate] (co3) [below=52mm of co2]{};
		\node [coordinate] (co4) [below=52mm of co1]{};
		\node [coordinate] (comid23) [below=26mm of co2]{};
		\node [coordinate] (comid14) [below=26mm of co1]{};
		\node [coordinate] (comid23r) [below=52mm of pmid23r]{};
		\node [coordinate] (comid14l) [below=52mm of pmid14l]{};
		\node (vc_input) [below=30mm of pmid23r,xshift=-5mm]{$\Vc$};
		\node (mode_input) [below=30mm of pmid14l,xshift=5mm]{$mode$};
		\draw [dashed] (co1)-- (co2) -- (co3) -- (co4)-- (co1){};
		\draw [dashed,->] (pmid23)-- (pmid23r) -- (comid23r)--(comid23){};
		\draw [dashed,->] (comid14)-- (comid14l) -- (pmid14l)-- (pmid14){};
	\end{tikzpicture}%
	}
	\vspace{-1em}%
	\caption{A hybrid automaton models the buck converter plant with hysteresis controller. 
	}%
	\figlabel{automaton_buckboost}%
	\vspace{-0.5em}%
\end{figure}%

\subsection{Candidate Invariant Checking Problem}
\seclabel{cicp}

The formal definition of the candidate invariant checking problem for CPS is described as follows.
\begin{definition}[Candidate Invariant Checking] Given a CPS model $\A$ with a set of candidate invariants $\hat{\Phi}$, $\tilde{\A}$ is a formal model converted from $\A$, a candidate invariant $\hat{\varphi} \in \hat{\Phi}$ is considered as an actually invariant property of $\tilde{\A}$ iff $Reach(\tilde{\A)} \models \hat{\varphi}$.
\deflabel{invariant checking}
\end{definition}

According to \defref{invariant checking}, if a CPS model $\A$ is a white box system that can be represented in terms of a formal model such as a CPIOA  $\tilde{\A}$, a hybrid system model checker may be used to check whether $\hat{\varphi}$ is an invariant property of $\tilde{\A}$. If there exists any reachable state of $\tilde{\A}$ that does not satisfy $\hat{\varphi}$, we can conclude that $\hat{\varphi}$ is not an actual invariant of the CPS model $\A$.


%% file: specification_mismatch.tex
\section{Cyber-Physical Specifications and Mismatches}
\seclabel{spec_mis}

In this section, we will formalize the concept of candidate cyber-physical specification mismatches of CPS, and introduce a potential method to detect such specification mismatches.
\subsection{Cyber-Physical Specifications}
\seclabel{cpspec}
Our goal is to find specifications that are invariants or conditional invariants, so we do not consider more general temporal logic formulas.
Under this assumption, a \emph{specification} is equivalent to a predicate over the state-space of the system.
Equivalently, a specification is a multi-sorted first-order logic (FOL) sentence (of a restricted class), so we assume the specification may be represented in the Satisfiability Modulo Theories (SMT) library standard language~\cite{demoura2009sbmf,smtlib}.
Under these assumptions, candidate invariants may be specified as quantifier-free SMT formulas over the variables of the SLSF model, where the type of a variable corresponds to the SMT sort.
For a formula $\phi$, let $\vars{\phi}$ be the set of variables appearing in $\phi$.
For a formula $\phi$:
\begin{enumerate*}[label=(\textit{\alph*})]
\item if $\vars{\phi}$ are all physical, then $\phi$ is a \emph{physical specification},
\item if $\vars{\phi}$ are all cyber, then $\phi$ is a \emph{cyber specification}, and
\item if $\vars{\phi}$ consists of both cyber and physical variables, then $\phi$ is a \emph{cyber-physical specification}.
\end{enumerate*}

Next, while we will try to infer interesting specifications $\phi$ using dynamic analysis later in the \papertype, we first highlight examples of specifications made a priori in system design, as these are necessary to define specification mismatches.
Let $\Sigma$ be a set of specifications for $\A$, which is a set of formulas over the variables of $\A$.
Referring to~\figref{hynger_status}, we separate the specification $\Sigma$ into sets of cyber and physical specifications, written respectively as $\specSetCyber$ and $\specSetPhysical$.
These specifications include assumptions about the physical environment, such as the value of gravitational force, temperature bounds, time constants, etc.
The physical specification also includes assumptions about the physical system's behavior and subcomponents, such as motor torque limits, temperature bounds of components, sampling rates, velocity limits, etc.
Here $\specSetCyber$ denotes the set of cyber specifications.
The cyber specifications include assumptions about software-physical interfaces, such as ADC resolution, DAC resolution, sampling rates, etc.
It also includes assumptions about the software system, subcomponents, and software-software interfaces, such as data formats, control flow, event orderings, etc.
For example, the buck converter has the following physical specifications:
\begin{align*}
%
\specPhysical{1} \ & \deq \ t \geq t_s \Rightarrow \Vout(t) = \Vref(t) \pm \Vripple, \\
\specPhysical{2} \ & \deq \ \Vs(t) = \Vs(0) \pm \delta_S, \\
\specPhysical{3} \ & \deq \ \Vref(t) = \Vref(0) \pm \delta_{ref},
%
%
\end{align*}
and $\specSetPhysical \ \deq \ \{\specPhysical{1}, \specPhysical{2}, \specPhysical{3} \}$.
Here, $\specPhysical{1}$ states that after some amount of constant startup time $t_s$, the output of the buck converter $\Vout(t)$ remains near a reference (desired) output voltage $\Vref(t)$.
Both $\specPhysical{2}$ and $\specPhysical{3}$ specify assumptions about the buck converter's environment, namely that its source voltage $\Vs$ and reference voltage $\Vref$ always remain near their initial values.
We note that while time may not typically be thought of as a state of the system, it can be encoded in this way easily, for example, by including a state variable $t$ with $\dot{t} = 1$.
To evaluate whether $\A$ has cyber-physical specification mismatches, we hypothesize that the cyber specification contains (at least a subset) of the physical specification.
This process is made more explicit in~\figref{hynger_status} and described next.
%
%
%
%
\begin{figure}[t!]%
	\centering%
	\includegraphics[width=0.8\columnwidth]{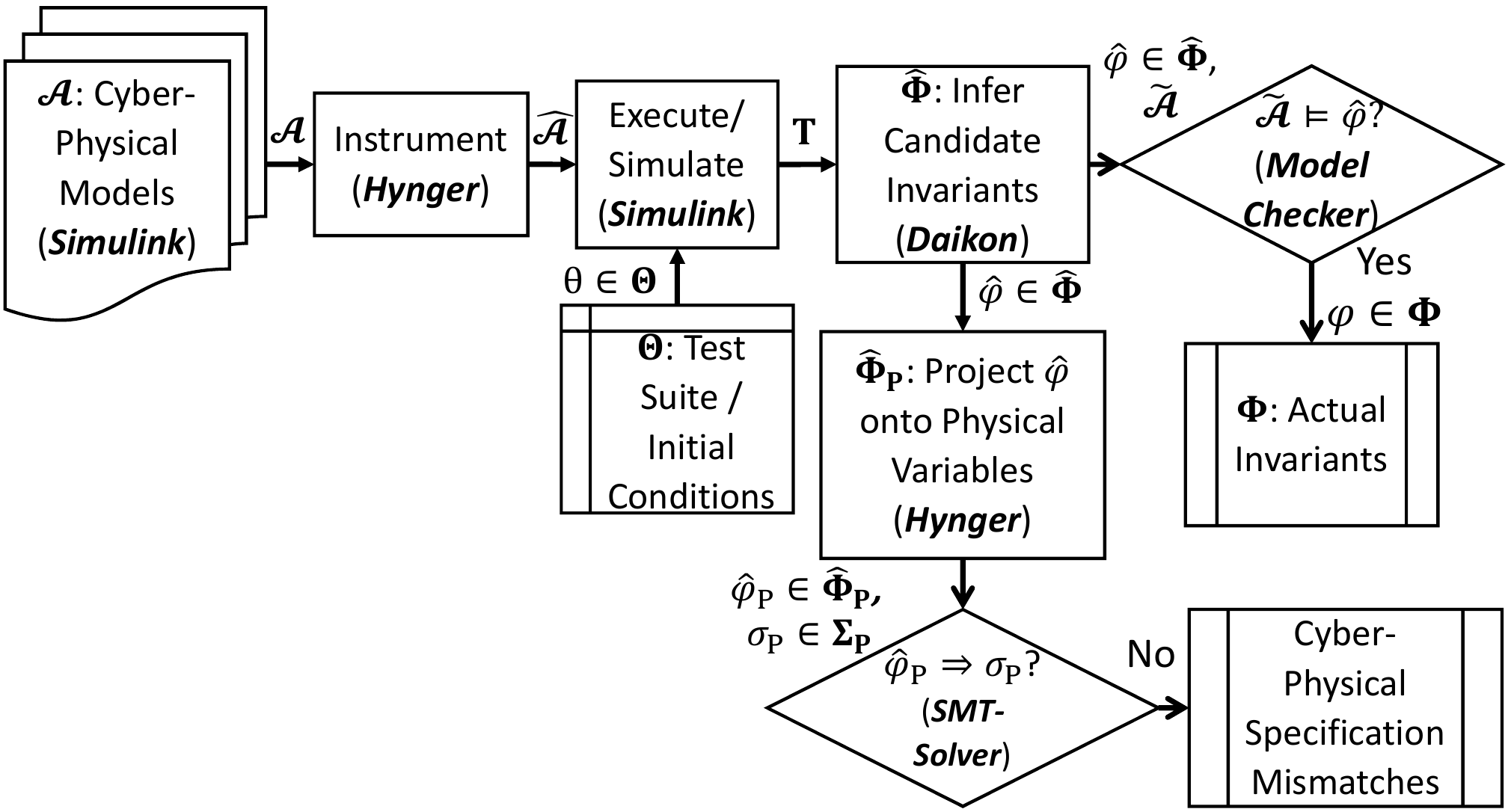}%
	\vspace{-0.5em}%
	\caption{Hynger overview, inference of physical specifications assumed by software, and cyber-physical specification mismatch identification.}%
	\figlabel{hynger_status}%
	\vspace{-1.5em}%
\end{figure}%
\subsection{Cyber-Physical Specification Mismatches}
A CPS model or implementation will be provided as an SLSF diagram, denoted $\A$ as formalized above.
Next, $\A$ is instrumented using the \toolhynger yielding a modified SLSF diagram $\hat{\A}$.
Now, $\hat{\A}$ is executed to generate a set of sampled, finite-precision traces $\mathrm{T}$ for each initial condition $\theta$ in a set of initial conditions $\Theta$, which effectively corresponds to a test suite.
The traces $\mathrm{T}$ are analyzed using dynamic analysis methods, such as Daikon, to generate a set of candidate invariants $\hat{\Phi}$, each element $\hat{\varphi}$ of which may be checked as actual invariants if $\A$ corresponds to a formal model (e.g., a CPIOA) or may be converted to one, $\tilde{\A}$.
If that is the case, then a hybrid system model checker may be employed to see if $\hat{\varphi}$ is an actual invariant $\varphi$, and the set of actual invariants $\Phi$ is collected.

\begin{definition}[Cyber-Physical Specification Mismatch]
Given an SLSF diagram $\A$ with a set of actual physical specifications $\specSetPhysical$, let $\hat{\Phi}_P \deq \hat{\Phi} \downarrow \HyngerVarSP$ be a set of candidate physical invariant, $\A$ has a cyber-physical specification mismatch iff: $\exists{\specPhysicalSymbol} \in \specSetPhysical$, $\forall{\hat{\varphi}_P} \in \hat{\Phi}_P$, $\specPhysicalSymbol \not\models \hat{\varphi}_P$.
\deflabel{cpsm}
\end{definition}
In \defref{cpsm}, $\hat{\Phi} \downarrow \HyngerVarSP$ denotes the projection or the restriction of $\hat{\Phi}$ to the set of software physical variable $\HyngerVarSP$.
In all cases, each candidate invariant $\hat{\varphi} \in \hat{\Phi}$ is projected (restricted) onto the software physical variables $\HyngerVarSP$ to yield a candidate physical invariant $\hat{\varphi}_P$ and corresponding set $\hat{\Phi}_P$.
Such a projection may be computed using quantifier elimination methods available in many modern SMT solvers, such as Z3~\cite{demoura2008tacas}\footnote{Z3 may be downloaded: \url{http://z3.codeplex.com/}.}.
Now, $\hat{\Phi}_P$ corresponds to the candidate, inferred physical invariants from the perspective of the cyber-physical system, each element of which may be compared to each element $\specPhysicalSymbol$ of a set of actual physical specifications $\specSetPhysical$.
Since $\hat{\varphi}_P$ and $\specPhysicalSymbol$ are both formulas, we construct new formulas $\hat{\varphi}_P \Rightarrow \specPhysicalSymbol$ and $\specPhysicalSymbol \Rightarrow \hat{\varphi}_P$, each of which may be discharged with an SMT solver.
If these checks are not valid, then these specifications are candidate \emph{cyber-physical mismatches}.
These checks basically compare whether the inferred specification and actual specification are more or less restrictive than one another, in terms of the sizes of corresponding sets of states satisfying the predicates.
We hypothesize that it is generally the case that the inferred physical specification should always be stronger than the actual physical specification, and only the check $\hat{\varphi}_P \Rightarrow \specPhysicalSymbol$ would be needed.
This would correspond to the case where the software's assumptions about the physical world are \emph{at least as} restrictive as those made in the actual physical specification.
For instance, suppose that the physical specification of the output voltage of the buck converter is $\specPhysicalSymbol \deq \ t \geq t_s \Rightarrow 4.8V \leq \Vout(t) \leq 5.2V$, and the candidate physical invariant is $\hat{\varphi}_P \deq \ t \geq t_s \Rightarrow 4.9V \leq \Vout(t) \leq 5.1V$, then the check of the formula $\hat{\varphi}_P \Rightarrow \specPhysicalSymbol$ using an SMT solver like Z3 will indicate that the system does not have a specification mismatch. 
Otherwise, if the candidate physical invariant is $\hat{\varphi}_P \deq \ t \geq t_s \Rightarrow 4.7V \leq \Vout(t) \leq 5.0V$, then the check of the formula $\hat{\varphi}_P \Rightarrow \specPhysicalSymbol$ will indicate that the system has a specification mismatch.
On the other hand, it may also be useful to check $\hat{\varphi}_P \Leftarrow \specPhysicalSymbol$, which would correspond to cases where the inferred physical specification is weaker than the actual physical specification.
In this case, there may be a trace that violates the actual specification, and this may be useful in analysis like falsification to drive simulations towards a violating behavior.
%
%

%% file: tool.tex
\section{Hynger: Generating Invariants for SLSF Models}
\seclabel{tool}
%
%
\toolhynger---HYbrid iNvariant GEneratoR---is a software tool developed for invariant inference of CPS models represented as SLSF block diagrams\footnote{A preliminary prototype of \toolhynger with examples is available online: \url{http://verivital.com/hynger/}.  The repository also includes Daikon input (\texttt{*.dtrace}) trace files generated from the examples, as well as the Daikon output candidate invariant (\texttt{*.inv}) files.}.
\toolhynger is written primarily in Matlab and uses the Matlab APIs to interact with SLSF diagrams.
\toolhynger also uses some Java code (natively inside Matlab) to interface with Daikon, which is written in Java.
Daikon versions 5.0.0 to 5.1.8 were tested with \toolhynger\footnote{Daikon may be downloaded: \url{http://plse.cs.washington.edu/daikon/}.}.

Given an SLSF model $\A$, \toolhynger automatically inserts callback functions into the model to print model variables at block inputs and outputs at certain events in the SLSF simulation loop. Consequently, a trace file generated by \toolhynger will then be formatted in the trace input format required by Daikon.
While configurable, the default behavior of \toolhynger is to add instrumentation (observation) points for every input and output signal for every block (recursively) in the SLSF diagram.
That is, \toolhynger walks the tree of blocks starting from the root, and for each $v \in \HyngerVertices$, adds instrumentation points for the input variables $\HyngerVarInput{v}$ and the output variables $\HyngerVarOutput{v}$ of $v$.
Of course, this may incur a drastic performance overhead, so if this is not desired, the user may select only a subset of the blocks to instrument and our performance results (see~\secref{experiment}) illustrate this distinction.
When an SLSF model is simulated with these instrumentation callback functions added by \toolhynger, it will generate a trace file in the input trace format for Daikon.
\toolhynger also provides the capability to automatically call Daikon from Matlab (by using an appropriate Java call to Daikon), which will then return the set of candidate invariants from each program point to the user.

%
%
%

The \toolhynger flow is summarized in~\figref{hynger_status}.
%
%
The inputs are:
\begin{enumerate*}[label=(\textit{\alph*})]
\item SLSF diagrams (containing embedded software code and a set of physical variables along with their physical dynamics models [e.g., ODEs]), and
\item a set of physical variables along with their dynamics models (specified as SLSF children diagrams), and
\item a test suite for the embedded software and initial conditions for the physical simulation (such as noisy initial conditions, $\theta \in \Theta$).
\end{enumerate*}
The output of the \toolhynger tool is a set of candidate invariants, which, when projected onto all the software physical variables $\HyngerVarSP$, represent a candidate specification the software assumes for the physical parts of the system.
Finally, candidate specifications can be checked for conformance with the actual physical requirements by comparing the two specifications: the actual physical specification and the candidate physical specification from the software perspective.

%
%
%
%


\todo{add simulink simulation loop image and description of how to define pre and post conditions}

\todo{add more case study details; fix microgrid setup as necessary; describe other case studies in more detail}

\todo{add more invariants synthesized, describe mismatch in more detail}

\todo{finish setup of z3 within matlab setup for hynger? using the spaceex converter setup?}

\begin{figure*}[!t]%
	\centering%
	\begin{lstlisting}[basicstyle=\ttfamily\scriptsize]
	/*@ requires n >= 0; // at least 0 elements
		@ requires \valid(b+ (0..n-1)); // all elements exist
		@ assigns \nothing; // no side effects
		@ ensures \result == \sum(0,n-1,\lambda integer j; b[j]);
		@ ensures \result >= 0; // false, array may be negative
	*/
	int sum_array(int b[], unsigned int n) {
			int i;
			int s = 0;
			/*@ loop invariant
					\forall integer j; (0 <= i <= n) ==> s == \sum(0,i-1,\lambda integer j; b[j]); */
			for (i = 0; i < n; i++) {
							s += b[i];
			}
			return s;
		}
	\end{lstlisting}
	\centering%
	\vspace{-1em}%
	\caption{Example C function that sums an array $\mathtt{b}$ of $\mathtt{n}$ integers.  Requirements on the function inputs (i.e., preconditions on $\mathtt{b}$ and $\mathtt{n}$ for the function to be called) are specified as $\mathtt{requires}$ assertions in the ACSL language.  Correctness specifications (i.e., postconditions following the function call) are specified as $\mathtt{ensures}$ assertions in the ACSL language.}%
	\figlabel{sum_array}%
\end{figure*}
\vspace{-1em}
\begin{figure*}[!ht]%
	\centering%
	\begin{lstlisting}[basicstyle=\ttfamily\scriptsize]
============== Precondition
..sum_array():::ENTER
b has only one value // it's a pointer to only one location of memory
b[] elements >= 0 // all elements were non-negative for this set of traces
n == 100 // all tests were 100 element arrays for this set of traces
size(b[]) == 100 // all tests were 100 element arrays
============== Postcondition
..sum_array():::EXIT
b[] == orig(b[]) // no side effects
return == sum(b[]) // does return the sum
sum(b[]) == sum(orig(b[]))
b[] elements >= 0
	\end{lstlisting}
	\centering%
	\vspace{-1em}%
	\caption{\tooldaikon candidate invariant output (with some additional markup in C-style comments for readability) for the $\mathtt{sum\_array}$ example from~\figref{sum_array}.}%
	\figlabel{sum_array_invariants}%
	\vspace{-2em}%
\end{figure*}

\subsection{Dynamic Invariant Inference with Daikon}
\seclabel{invariants}
Next, we illustrate the dynamic invariant inference methodology used by Daikon on a pure software example.
However, this pure software example (a C function) is actually specified for the controller in the buck converter case study (shown in \figref{case_study_overview}) in a different manner.
The loop in the controller SLSF model of~\figref{case_study_controller_hysteresis} also computes a sum of an array, and Daikon can find this specification for both the SLSF controller model using \toolhynger, and the C-frontend for the following example. Note that, in ~\figref{case_study_controller_hysteresis} the digitized output voltage from the buck-converter plant is used to determine the mode of the switch. Here, $\Vtol$ is denoted by the variable $\mathtt{Vtol}$, $\Vref$ is $\mathtt{Vref}$. We highlight that the controller computes a moving average by summing an array.  With \toolhynger and Daikon, we automatically infer that the result of this is the sum of the samples, similar to the sum return specification shown in~\figref{sum_array_invariants} found for the C function in~\figref{sum_array}.
\vspace{-0.5em}
\paragraph*{Example C Program, Formal Specification, and Candidate Invariants Inferred}
\figref{sum_array} shows an example C function to illustrate the use of dynamic analysis with \tooldaikon to find candidate invariants.
The function computes and returns the sum of an array of integers.
This example was recreated from an example in the original \tooldaikon paper~\cite{ernst2001tse}.
Additionally, a formalized correctness specification is given in the modern ANSI/ISO C Specification Language (ACSL), used by tools such as \toolframac~\cite{cuoq2012sefm}.
Using \tooldaikon and a small suite of unit tests, we were able to successfully find the invariant that returns from the function $\mathtt{sum\_array}$, the returned value is the sum of the elements in the array $\mathtt{b}$.
The suite of tests included arrays with:
%
\begin{enumerate*}[label=(\textit{\alph*})]
\setlength\itemsep{0.1em}
\item all the same length and same elements,
\item all the same length and uniformly randomly chosen elements,
\item different lengths and all the same elements, and
\item different lengths and uniformly randomly chosen elements.
\end{enumerate*}
\tooldaikon successfully found the sum postcondition in all these cases with only a few test conditions.
%
%
The candidate invariant outputs of \tooldaikon appear in~\figref{sum_array_invariants}, where we can see Daikon has inferred a candidate invariant that the function returns the sum of an array.
We highlight that we find the sum return result of the moving average filter from~\figref{case_study_controller_hysteresis} using \toolhynger and Daikon.

%
%

%% file: experiment.tex
\section{Experimental Results}
\seclabel{experiment}
\toolhynger was tested on Windows 10 64-bit using Matlab 2016b, and 2017a, executed on a x86-64 laptop with a 2.3 GHz dual-core Intel i5-6200U processor and 12 GB RAM.
All performance metrics reported were recorded on this system using Matlab 2017a.
We tested and evaluated \toolhynger using a number of SLSF examples, including:
\begin{enumerate*}[label=\textit{(\alph*)}]
\item the closed-loop buck converter with sensor and hysteresis controller described in~\secref{buck_experimental_result} and detailed further in~\cite{nguyen2014arch},
\item a solar array case study that uses a buck-boost converter~\cite{nguyen2014tec},
\item benchmarks from S-TaLiRo~\cite{annpureddy2011tacas},
\item benchmarks from Breach~\cite{donze2010cav,jin2013hscc},
\item benchmarks created as a part of the ARCH 2014 CPSWeek workshop (particularly~\cite{nguyen2014arch,jin2014arch}) and
\item example models provided by Mathworks.
\end{enumerate*}
Overall, these examples vary from fairly simple with tens of blocks (such as the buck converter case study we detail), to complex (with hundreds of blocks).
%
%
\vspace{-0.5em}
\paragraph*{Runtime Overhead from Instrumentation with \toolhynger and Invariant Inference with Daikon}
%
%
First, we present an aggregate performance evaluation for some of these examples in~\tabref{hynger_overhead}, with column descriptions appearing in the caption.
Overall, the performance overhead of instrumenting diagrams and performing invariant inference is around an order of magnitude increase in the best cases, and two-to-three orders of magnitude increase in the worst cases, which we note is comparable with typical Daikon instrumentation frontends like Valgrind's overhead~\cite{ernst2007scp,nethercote2007pldi}.
We conducted performance profiling of \toolhynger and identified the main source of overhead (about $75$ to $90$ percent) as file I/O operations.
%
Additionally, as \toolhynger has several different usage scenarios and operating modes (where it may be used to instrument few blocks [subsystem and function blocks by default], many blocks [all blocks except ones such as constants, scopes, etc.], every single block, or user-selected blocks), the table illustrates these differences to give some comparison of how the methods scale on a given model.
Next, we will describe two CPS case studies in details to evaluate the capability of \toolhynger in detecting cyber-physical specification mismatches.
The first model is the closed-loop buck converter that has been used to illustrate the concepts of this paper, and the second model is derived from a collection of the automotive powertrain control benchmarks proposed by Toyota~\cite{jin2013hscc}.

\begin{table*}[!t]
	\centering
	\resizebox{\textwidth}{!}{
	\small{
		\begin{tabular}{l|l|l|l|l|l|l|l|l|l|l}
Model & Solver & Tmax & Sim & SimInst & Inv & Overhead &  BDAll & BDInst & BDPct\\
\hline
buck (\secref{buck_experimental_result}) & ode45 & $0.0083$ & $6.2985$ & $38.4518$ & $5.7335$ & $7.0152$ & $14$ & $3$ & $21.4286$ \\
buck (\secref{buck_experimental_result}) & ode45 & $0.0083$ & $6.4567$ & $44.698$ & $7.0913$ & $8.021$ & $14$ & $4$ & $28.5714$ \\
buck (\secref{buck_experimental_result}) & ode45 & $0.0083$ & $6.5301$ & $78.3176$ & $7.2224$ & $13.0993$ & $14$ & $14$ & $100$ \\
heat25830~\cite{annpureddy2011tacas} & ode45 & $50$ & $4.6913$ & $254.5776$ & $14.09$ & $57.2692$ & $28$ & $1$ & $3.5714$ \\
heat25830~\cite{annpureddy2011tacas} & ode45 & $50$ & $4.7328$ & $2882.7808$ & $15.6488$ & $612.4233$ & $28$ & $10$ & $35.7143$ \\
fuel1~\cite{jin2014arch} & ode15s & $15$ & $5.3747$ & $976.6274$ & $7.923$ & $183.182$ & $208$ & $17$ & $8.1731$ \\
fuel1~\cite{jin2014arch} & ode15s & $15$ & $4.2131$ & $2824.2804$ & $11.604$ & $673.1137$ & $208$ & $63$ & $30.2885$ \\
fuel2~\cite{jin2014arch} & ode15s & $20$ & $3.3838$ & $36.8312$ & $2.9881$ & $11.7674$ & $25$ & $6$ & $24$ \\
fuel2~\cite{jin2014arch} & ode15s & $20$ & $2.7353$ & $42.4074$ & $3.2771$ & $16.7018$ & $25$ & $13$ & $52$ \\
fuel3~\cite{fan15_c2e2_benchmark} & ode15s & $20$ & $3.7425$ & $292.9976$ & $4.1131$ & $79.3892$ & $90$ & $11$ & $12.2222$ \\
fuel3~\cite{fan15_c2e2_benchmark} & ode15s & $20$ & $3.6083$ & $945.3992$ & $4.3904$ & $263.2236$ & $90$ & $46$ & $51.1111$ \\
		\end{tabular}
	}}
	\caption{\toolhynger performance results for several of the examples evaluated.  Solver is the ODE solver used by SLSF.  Tmax is the virtual simulation time in seconds (i.e., time from the perspective of the model).  All runtime results are in seconds and are the mean of $20$ runs.  Sim is the simulation runtime ($s$).  
Inv is the invariant generation runtime (Daikon) ($s$).  Overhead is the overall relative performance overhead (extra runtime) ($\times$) using \toolhynger and Daikon versus only SLSF simulation (i.e., $((SimInst + Inv) / Sim)$).  BDInst and BDAll are the numbers of block diagrams instrumented and the overall number of block diagrams, respectively.  BDPct is the percentage ($\%$) of block diagrams instrumented using different \toolhynger modes of operation (i.e., $BDInst / BDAll$).}%
	\tablabel{hynger_overhead}%
\vspace{-3em}%
\end{table*}
\vspace{-0.5em}
\subsection{Closed-Loop Buck Converter Cyber-Physical Specification Mismatch}
\seclabel{buck_experimental_result}
%
%
\todo{expand and give additional details: define the actual physical vs. inferred physical specifications and sets thereof: walk through the diagrammatic process with every step for buck-boost}
A basic cyber-physical specification mismatch is easy to encode in the buck converter, since the software controller inherently uses a tolerance to encode the desired output voltage ripple.
This hysteresis tolerance band is typically chosen based on the system dynamics and desired output voltage ripple to ensure the output voltage meets the ripple specification.
As a concrete example, the physical specification may contain a fixed constraint that $\Vout = \Vref \pm \Vripple$, e.g., $\Vref = 5V$ and $\Vripple = 0.1V$.
The hysteresis band $\Vtol$ is then selected based on the system dynamics to ensure $4.9V \leq \Vout \leq 5.1V$ so that it meets the requirements of the physical specifications defined by $\specSetPhysical$ in~\secref{cpspec}.

\paragraph*{Sources of Cyber-Physical Specification Mismatches of the Closed-Loop Buck Converter} There are different possibilities of specification mismatch that may occur to the closed-loop buck converter.
We present three scenarios that result in specification mismatches.
First, if the plant parameters change (i.e., different circuit elements are used), and the software is not updated with a new hysteresis band $\Vtol$ to accommodate the changes in the plant dynamics, then a specification mismatch manifests.
This mismatch can be detected using \toolhynger and the methodology described in this \papertype.
%
Of course, this is a somewhat obvious mismatch, as the controller relies on variables computed as functions of the plant parameters (here, the $R$, $L$, and $C$ values, as well as the source and desired/reference output voltage values). So if these plant components are changed, clearly the software must be updated.
Second, the hysteresis controller is initially constructed using wrong information about the physical evolution of the plant.
In fact, the hysteresis band $\Vtol$ is far different from the actual output voltage ripples $\Vripple$ of the plant.
Third, the analog sensor of the buck converter may have ADC conversion errors that reduce the accuracy of the voltage measurement. These errors can be an offset error, a full-scale error, differential and integral non-linearity errors, etc.
Moreover, a typical error that cannot be avoided in ADC sensor is the quantization error~\cite{staller2005}. Overall, these conversion errors may cause a significant impact to result in system failures.
%
%
%
\todo{more details on how Hynger finds this}
\begin{figure}[t!]%
	\centering%
	\includegraphics[width=0.8\columnwidth]{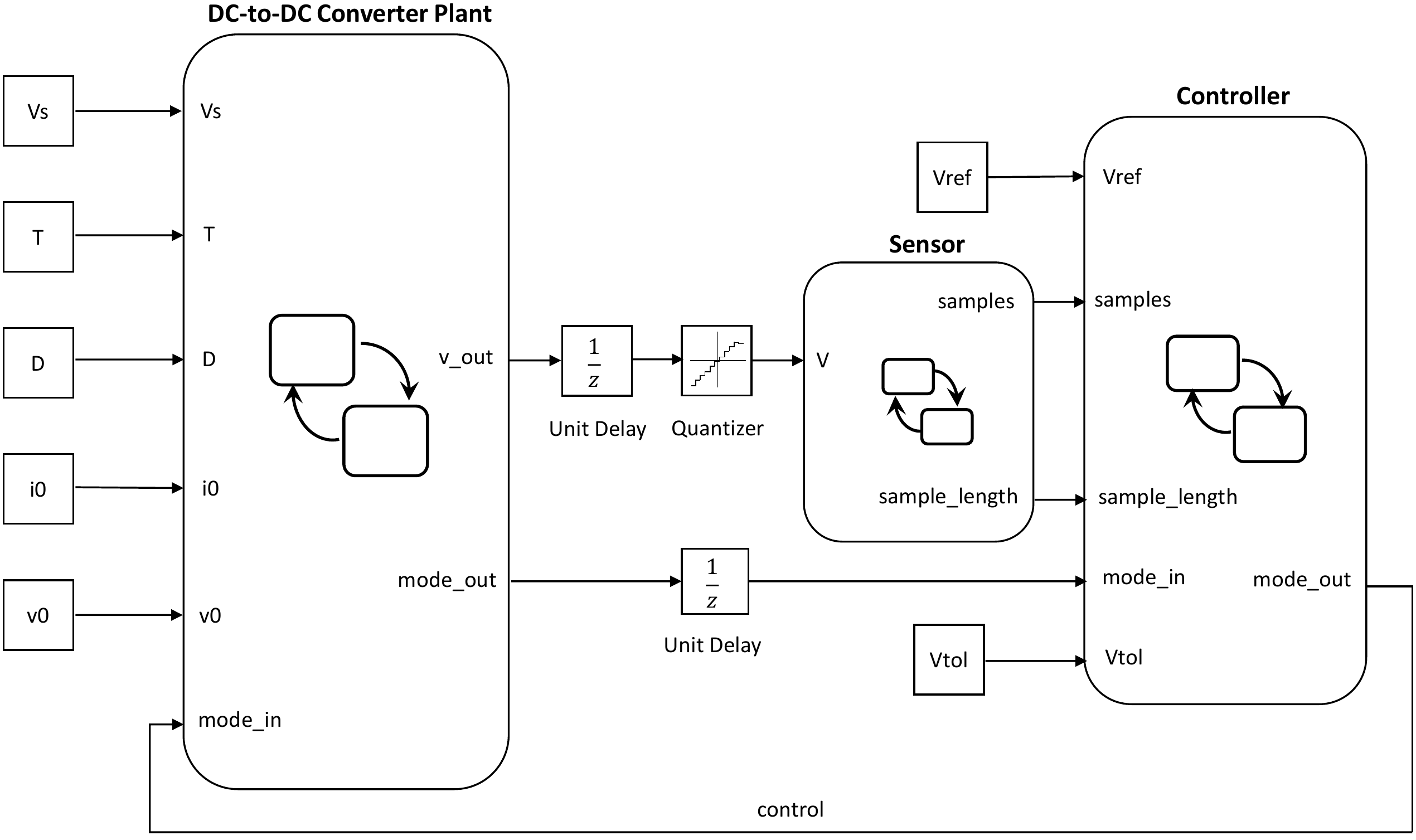}%
	\caption{General CPS case study architecture overview of the buck converter in SLSF. The system is composed of a plant (physical system) model, a controller (software/cyber), and potentially sensor and actuator models. The cyber model uses some of the physical model output states to determine a control action or input. The controller in SLSF appears in~\figref{case_study_controller_hysteresis}, and the sensor model appears in~\figref{case_study_sensor}. An example of this closed-loop buck converter including only plant and controller can be formally represented as the hybrid automaton in~\figref{automaton_buckboost}.}%
	\figlabel{case_study_overview}%
	\vspace{-1em}
\end{figure}%
%
%
%
\begin{figure}
\centering
\begin{minipage}{.5\textwidth}
  \centering
  \includegraphics[width=0.9\linewidth, height = 46mm]{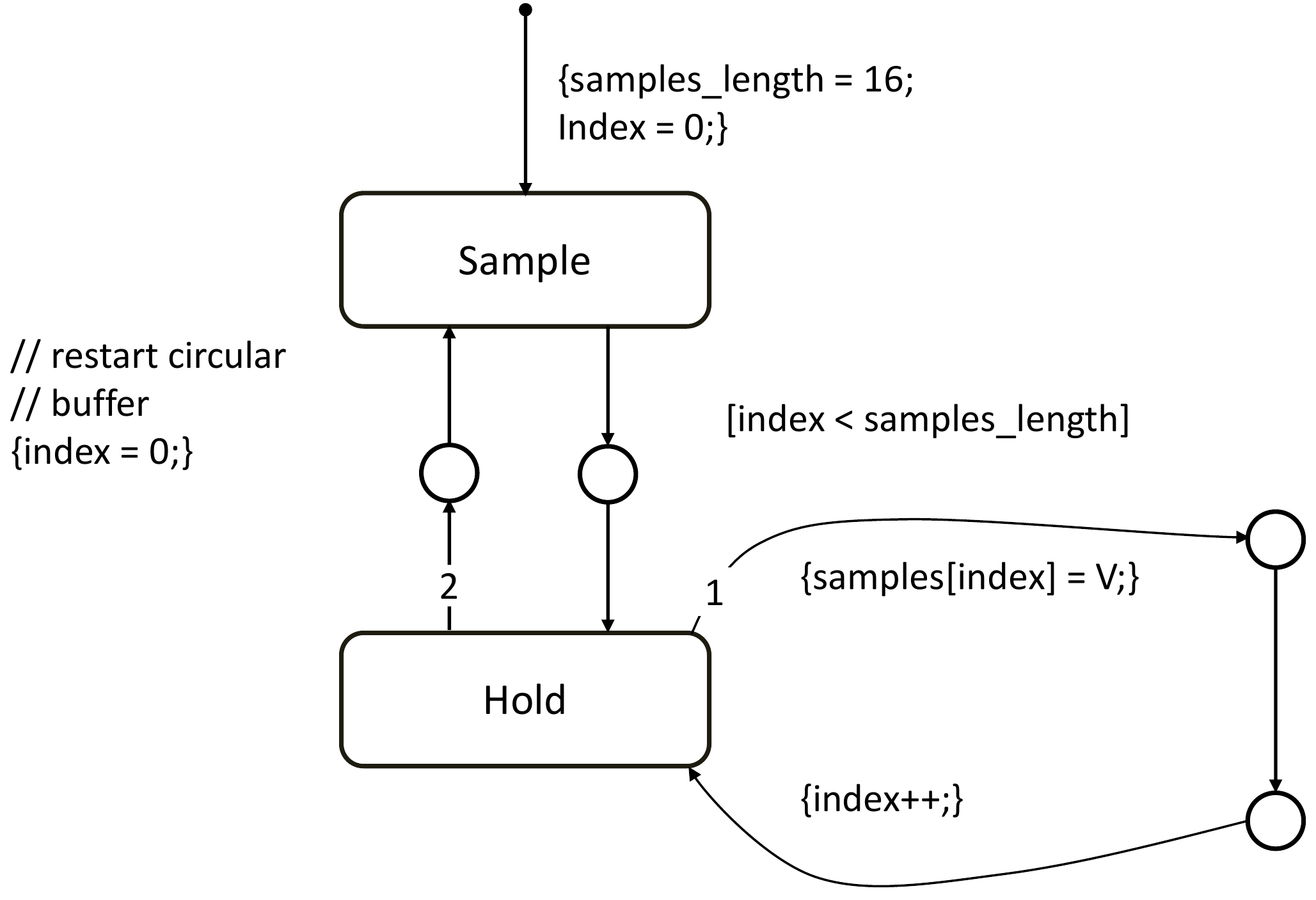}
	\captionsetup{width=0.9\linewidth}
  \captionof{figure}{Stateflow model of sensor with a sample and hold for the buck converter case study.}
  \figlabel{case_study_sensor}
\end{minipage}%
\begin{minipage}{.5\textwidth}
  \centering
  \includegraphics[width=0.9\linewidth, height = 47mm]{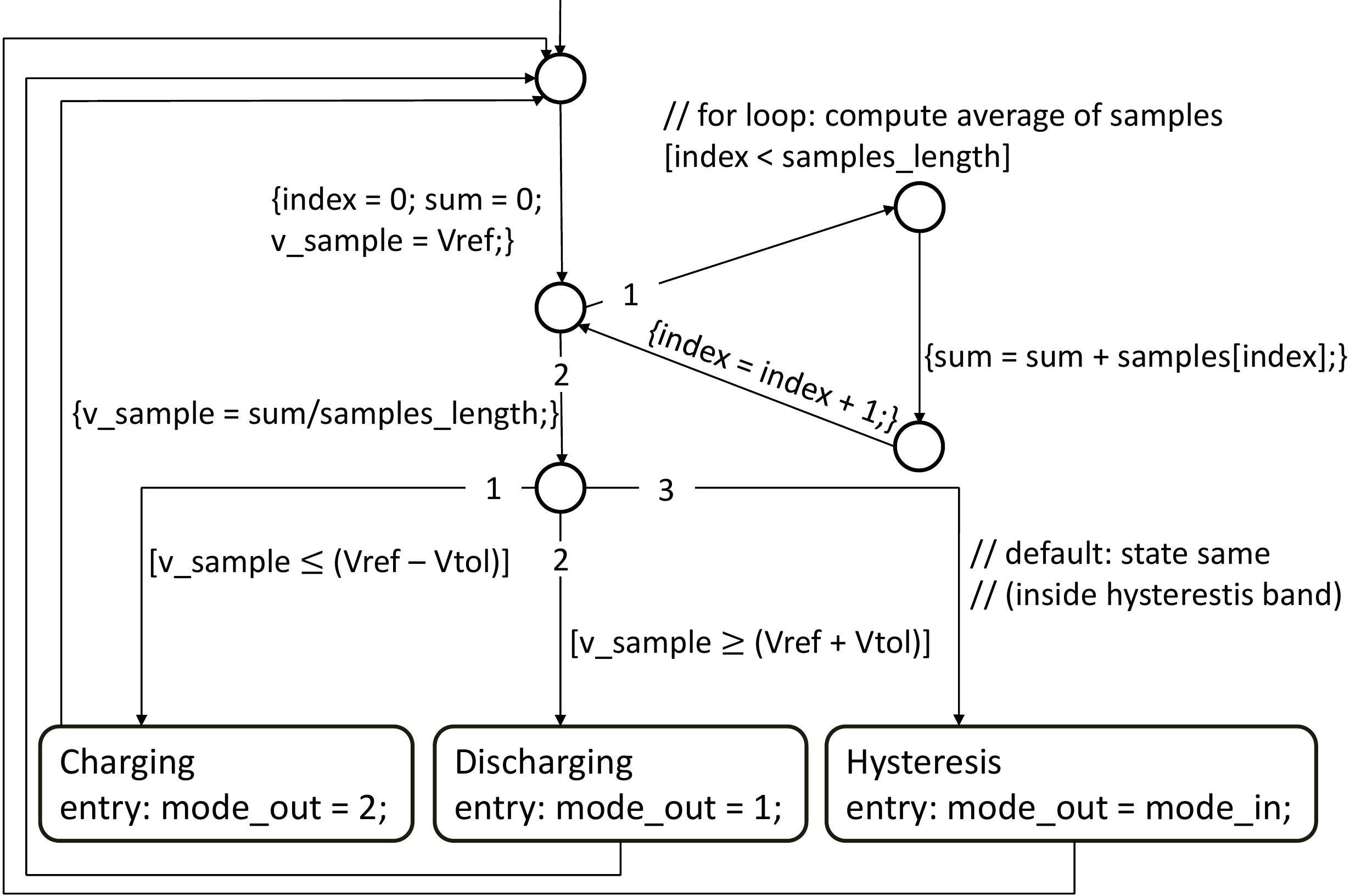}
	\captionsetup{width=0.9\linewidth}
  \captionof{figure}{Stateflow model of the buck-converter voltage hysteresis controller.}
  \figlabel{case_study_controller_hysteresis}
\end{minipage}
\end{figure}


\paragraph*{Experimental Results in Identifying Cyber-Physical Specification Mismatches of the Closed-Loop Buck Converter}
We consider the closed-loop buck converter $\A$ shown in \figref{case_study_overview} with $\Vs = 100$, $\Vref = 48V$, $\Vripple = 5\%\Vref = 2.4V$, and assume that $\delta_S$, $\delta_{ref}$ are equal to zero. The physical specification of the output voltage is $\specPhysicalSymbol \deq \ t \geq t_s \Rightarrow 45.6V \leq \Vout(t) \leq 50.4V$.
%
%
%
For the initial setup, with $R = 6 \Omega$, $L = 2.65 mH$, $C = 2.2 mF$, and a sampling frequency $f_s = 60 kHz$, the magnitude bound of the output voltage inferred from \toolhynger and Daikon is $\hat{\varphi}_P \deq \ t \geq t_s \Rightarrow 46.559V \leq \Vout(t) \leq 50.203V$.
Then, $\hat{\varphi}_P$ is considered as the candidate invariant of the system since the formula $\hat{\varphi}_P \Rightarrow \specPhysicalSymbol$ is true.
%
%
Next, we investigate different possibilities of cyber-physical specification mismatches that may occur when changing the source voltage, the desired/reference output voltage, the sampling frequency, and the plant parameters of the buck converter.

First, we increase the source voltage $\Vs$ from $100V$ to $120V$, the new magnitude bound of the output voltage inferred from \toolhynger and Daikon is $\hat{\varphi}_P \deq \ t \geq t_s \Rightarrow 46.804V \leq \Vout(t) \leq 51.118V$. Then, the formula $\hat{\varphi}_P \Rightarrow \specPhysicalSymbol$ is false, that indicates the system may have a cyber-physical specification mismatch.

Second, we drop the desired/reference output voltage $\Vref$ to $36V$. Thus, the physical specification of the output voltage becomes $\specPhysicalSymbol' \deq \ t \geq t_s \Rightarrow 34.2V \leq \Vout(t) \leq 37.8V$. In this case, the inferred physical specification of the output voltage from \toolhynger and Daikon becomes $\hat{\varphi}_P' \deq \ t \geq t_s \Rightarrow 35.068V \leq \Vout(t) \leq 39.053V$, so that the formula $\hat{\varphi}_P' \Rightarrow \specPhysicalSymbol'$ is false. Therefore, changing the reference output voltage may also produce a cyber-physical specification mismatch for the buck converter.

Third, we decrease the sampling frequency $f_s$ from $60kHz$ to $30kHz$. As a result, the new inferred physical specification of the output voltage from \toolhynger and Daikon is $\hat{\varphi}_P \deq \ t \geq t_s \Rightarrow 45.853V \leq \Vout(t) \leq 51.091V$. The check of the formula $\hat{\varphi}_P \Rightarrow \specPhysicalSymbol$ will return false to indicate that the system may contain a cyber-physical specification mismatch.

Next, we keep the controller unchanged and vary the values of $R$, $L$, and $C$ to change the plant parameters.
We then run the buck converter with \toolhynger in conjunction with Daikon, and collect candidate physical specifications associated with the output voltage.
The comparison between the actual physical specification $\specPhysicalSymbol$ and the physical specification $\hat{\varphi}_P$ inferred from \toolhynger and Daikon is shown in \tabref{buck_result}, and also illustrated in \figref{buck_vc_time}.
Note that in \tabref{buck_result}, $\hat{\varphi}_P$ describes the magnitude bound of the output voltage when $\ t \geq t_s$.
The checks of the formula $\hat{\varphi}_P \Rightarrow \specPhysicalSymbol$ occasionally return $False$, that are depicted in \figref{buck_vc_time} when the bound of the inferred output voltage overlaps its actual bound.
This indicates that changing the plant parameters without updating the controller may produce cyber-physical specification mismatches.
That also proves the capability of \toolhynger and our proposed methodology in automatically detecting a candidate cyber-physical specification mismatch of CPS.
%

%

%
Another possibility of the specification mismatch may occur when the controller is encoded based on wrong information about the plant. For the buck converter, the hysteresis controller is built with an assumption that the output voltage ripple $\Vripple $ is equal to $5\%$ of the reference voltage $\Vref$.
However, the actual value of $\Vripple$ may be much smaller than this assumption percentage. The percentage of the output voltage ripple of the buck converter is calculated as follows~\cite{erickson2004book},
\begin{align}
\vspace{-1em}
\frac{\Vripple}{\Vref} & = \frac{1 - D}{8LCf_s^2}\eqlabel{Vripple},
\vspace{-1em}
\end{align}
where $D = \frac{\Vref}{\eta\V_S}$ is a duty cycle, and $\eta$ is an efficiency coefficient of the converter. Here, with $L = 2.65 mH$, $C = 2.2 mF$, $f_s = 60kHz$, $\eta = 0.79$, $\Vref = 48V$, and $\V_S = 100V$, the percentage of the output voltage ripple is approximately equal to $0.0002\%$.
Thus, the hypothesized output voltage ripple used to build the controller is far larger than the actual output voltage ripple calculated by \eqref{Vripple}.
It definitely shows that the system may have specification mismatches since the controller is encoded depending on wrong information about the physical plant.


Furthermore, changing the length of voltage measurement array (samples$\_$length) in the sensor of the buck converter (shown in~\figref{case_study_sensor}) may also cause a specification mismatch. For example, if we increase it from $16$ to $32$, the inferred physical specification using \toolhynger and Daikon becomes $\hat{\varphi}_P \deq \ t \geq t_s \Rightarrow 46.095V \leq \Vout(t) \leq 50.788V$, which no longer implies the actual physical specification of the output voltage $\specPhysicalSymbol \deq \ t \geq t_s \Rightarrow 45.6V \leq \Vout(t) \leq 50.4V$.
%
%
%
%
\begin{table*}[!t]
	\centering
	\small{

		\begin{tabular}{|c|c|c|c|}
		\hline
		Parameter Values & $\hat{\varphi}_P$ & $\hat{\varphi}_P \Rightarrow \specPhysicalSymbol$ & $\specPhysicalSymbol \Rightarrow \hat{\varphi}_P$ \\ \hline
		$R = 4  \Omega$, $L = 2.65 mH$, $C = 2.2 mF$&$45.137V \leq \Vout(t) \leq 49.723V$&$False$&$False$ \\ \hline
		$R = 8 \Omega$, $L = 2.65 mH$, $C = 2.2 mF$&$46.964V \leq \Vout(t) \leq 50.405V$&$False$&$False$ \\ \hline
		$R = 6 \Omega$, $L = 0.65 mH$, $C = 2.2 mF$ &$47.141V \leq \Vout(t) \leq 50.074V$&$True$&$False$ \\ \hline
		$R = 6 \Omega$, $L = 6.65 mH$, $C = 2.2 mF$&$45.429V \leq \Vout(t) \leq 50.439V$&$False$&$True$ \\ \hline
		$R = 6 \Omega$, $L = 2.65 mH$, $C = 1.2 mF$&$45.426V \leq \Vout(t) \leq 51.109V$&$False$&$True$ \\ \hline
		$R = 6 \Omega$, $L = 2.65 mH$, $C = 3.2 mF$&$46.859V \leq \Vout(t) \leq 49.774V$&$True$&$False$ \\ \hline
		\end{tabular}}
		\caption{Experimental data showing the comparison between actual physical specifications and inferred physical invariants from \toolhynger and Daikon of the buck converter system. Here, the  plant component is changed due to the changes of $R$, $L$, and $C$ values.}
	\tablabel{buck_result}%
	\vspace{-2em}%
\end{table*}
%
%
%
%
%
%
\begin{figure}[!ht]%
	\centering%
	\subfloat{
	\includegraphics[width=0.9\columnwidth]{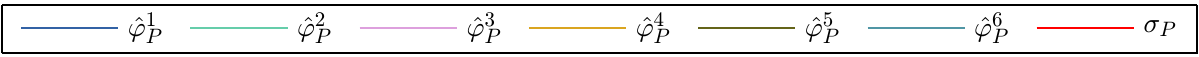}%
	}
		
	\vspace{-1em}
	\subfloat{
	\includegraphics[width=0.6\columnwidth]{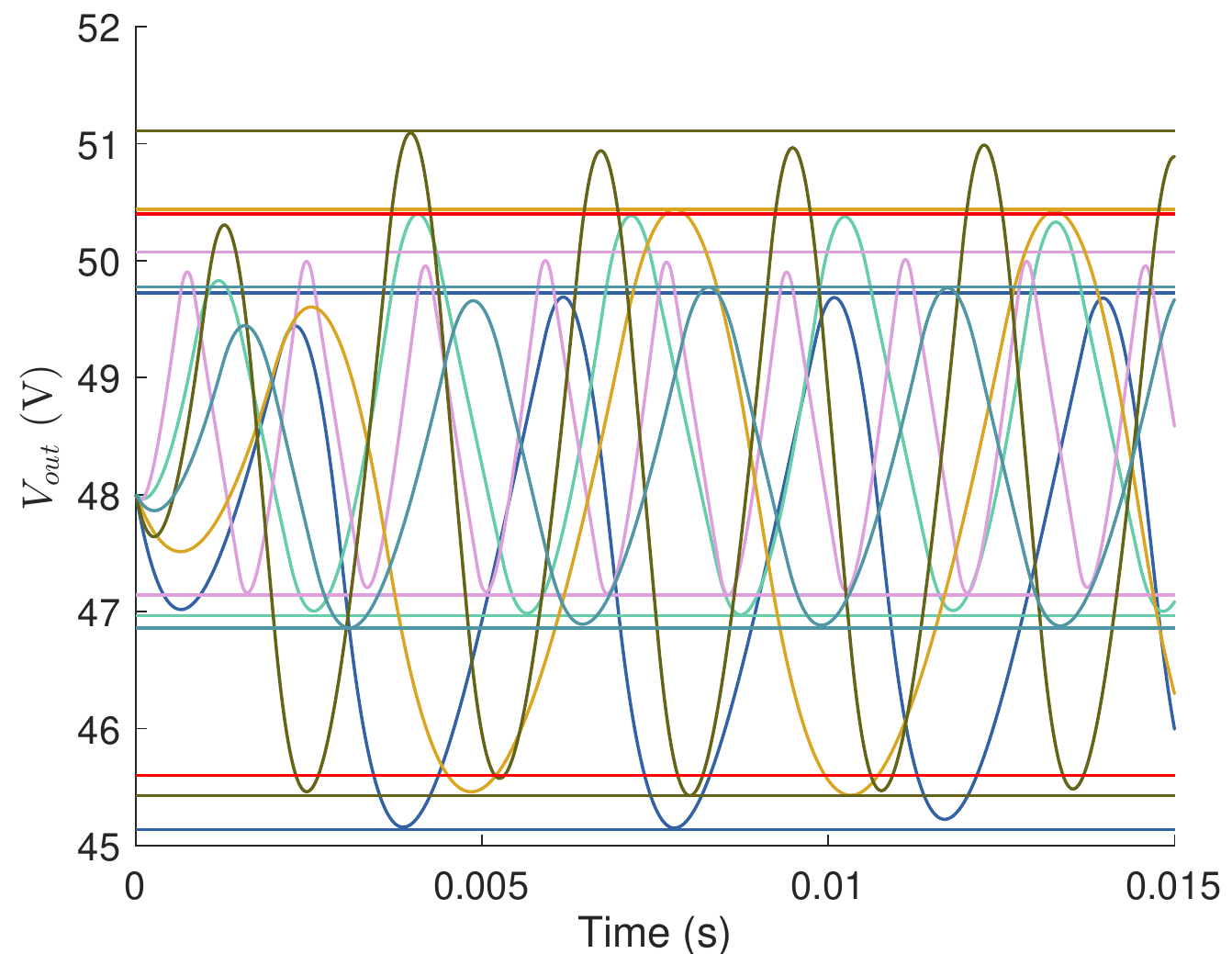}
	}
	\caption{A plot represents simulation traces and magnitude bounds of $\Vout$ of the buck converter with different values of $R$, $L$, and $C$. Here, $\specPhysicalSymbol$ denotes the actual bound of $\Vout$, and $\inferredSpecPhysical{k}$, $k\in[1,6]$ denotes the inferred bound of $\Vout$ listed orderly in \tabref{buck_result}. }%
	\figlabel{buck_vc_time}%
	\vspace{-1em}
\end{figure}%

\subsection{Abstract Fuel Control System Benchmarks}
\seclabel{powertrain}
In the second case study, we present the potential cyber-physical specification mismatches of the abstract fuel control (AFC) system benchmarks provided by Toyota~\cite{jin2013hscc,jin2014arch}, and further studied in \cite{fan15_c2e2_benchmark}.
The goal of these benchmarks is to determine the fuel rate that should be injected into the manifold to maintain the air-fuel ratio within a desirable range using the feedforward and Proportional-Integral (PI) controllers.
Particularly, we focus on the third model of the benchmarks including a sequence of Simulink blocks and Stateflow chart that increase levels of sophistication and fidelity of the system~\cite{fan15_c2e2_benchmark}.
The model consists of four operation modes and four continuous variables. The modes include \emph{startup}, \emph{normal}, \emph{power}, and \emph{failure}; and the variables are
\begin{enumerate*}[label=(\textit{\alph*})]
\item $p$: an intake manifold pressure,
\item $p_e$: an intake manifold pressure estimate,
\item $\lambda$: an air-fuel ratio, and
\item $i$: an integrator state, PI control signal.
\end{enumerate*}
The evolution of the continuous variables in each mode is governed by nonlinear polynomial differential equations as follows,
\begin{align}
%
\dot{p} & = c_1(2\theta(c_{20}p^{2} + c_{21}p + c_{22}) - \dot{m_c})\eqlabel{pressure}\\
\dot{p_e} & = c_1(2c_{23}\theta(c_{20}p^{2} + c_{21}p + c_{22}) - (c_2 +c_3\omega p_e + c_4\omega p_e^2 +c_5\omega^2p_e))\eqlabel{pressureE}\\
\dot{\lambda} & = c_{26}(c_{15} + (c_{16}c_{25}F_c + c_{17}c_{25}^2F_c^2 + c_{18}\dot{m_c} + c_{19}\dot{m_c}c_{25}F_c - \lambda)\eqlabel{AFrate}\\
\dot{i} & = c_{14}(c_{24}\lambda - c_{11})\eqlabel{PIcontrol},
\end{align}
where $F_c = \frac{1}{c_{11}}(1 + i +c_{13}(c_{24}\lambda - c_{11}))(c_2 +c_3\omega p_e + c_4\omega p_e^2 +c_5\omega^2p_e)$, and $\dot{m_c} = c_{12}(c_2 +c_3\omega p + c_4\omega p^2 +c_5\omega^2p)$. $\theta$ and $\omega$ are throttle angle (in degrees) and engine speed inputs (in $rpm$), respectively. The values of all constant parameters $c_j, j\in[1,25]$, $\theta$ and $\omega$ are specified in \cite{jin2013hscc}.
We note that this system can be formally represented as a closed-loop CPIOA, which is the parallel composition of a plant and controller model, and both of them have three exogenous inputs including $\theta$, $\omega$, and sensor failure event \emph{fail\_event}~\cite{fan15_c2e2_benchmark}.
\paragraph*{AFC Plant Model} The plant can be modeled as a CPIOA with a single mode and two output physical variables $p$, $\lambda$ whose continuous evolutions over time are described in \eqref{pressure} and \eqref{AFrate}, respectively. This model has an input cyber variable $F_c$, that is a fuel command.
\paragraph*{AFC Controller Model} The controller model is a CPIOA with four operation modes including \emph{startup}, \emph{normal}, \emph{power}, and \emph{failure}.  The controller has two output physical variables $p_e$, and $i$ whose continuous evolutions over time are described in \eqref{pressureE} and \eqref{PIcontrol}, respectively. Here, $p$ and $\lambda$ are considered as two input cyber variables of the controller.

Reachability analysis of a sophisticated system like the AFC system is a major contribution to both industrial and research community.
However, it is a challenge to design and verify such a system using existing hybrid system verification tools. Instead, we can attempt to verify some safety requirements of the system.
The AFC system has several actual physical specifications that can be found in~\cite{duggirala2015}. In this section, we select two main physical specifications to evaluate the capability of \toolhynger and the proposed methodology. 
%
The first physical specification requires the undershoot and overshoot of the air-fuel ratio of the system should be in the settling region of $\pm2\%$ of its reference value $\AFref$.
The second physical specification requires the air-fuel ratio should be maintained within $\pm2\%$ of $\AFref$ in the $normal$ mode when $t \geq t_s$.
These properties can be formally expressed as: 
\begin{align}
	\specPhysical{1} & \deq \ mode = startup \wedge t \leq t_s \Rightarrow 0.98\AFref \leq \lambda(t) \leq 1.02\AFref \eqlabel{fuel_start} \\
	\specPhysical{2} & \deq \ mode = normal \wedge t \geq t_s \Rightarrow 0.98\AFref \leq \lambda(t) \leq 1.02\AFref \eqlabel{fuel_normal}.
\end{align}
Initially, we set $\AFref = 14.7$, $\theta \in [8.8^{\circ}, 90^{\circ}]$, $w = 1800 rpm$ $t_s = 9.5s$, and the maximum simulation time $T_{max} = 20s$, the proportional and integral gains of the PI controller are $c_{13} = 0.04$ and $c_{14} = 0.14$, respectively.
Next, we investigate different possibilities of cyber-physical specification mismatches for each physical specification. For the first physical specification $\specPhysical{1}$, the AFC system may have specification mismatches when changing the engine speed and throttle inputs.
For the second physical specification $\specPhysical{2}$, the system may contain specification mismatches when changing controller and plant parameters.

\paragraph*{Cyber-physical specification mismatches according to $\specPhysical{1}$}
With the initial setup mentioned earlier, the physical specification in \eqref{fuel_start} becomes $\specPhysicalSymbol \deq \ mode = startup \wedge t \leq 9.5 \Rightarrow 14.406 \leq \lambda(t) \leq 14.994$.
Here, the magnitude bound of the air-fuel ratio at the $startup$ mode of the system inferred from \toolhynger and Daikon is $\inferredSpecPhysical{1}  \deq \ mode = startup \wedge t \leq 9.5 \Rightarrow 14.505 \leq \lambda(t) \leq 14.97$.
Thus, the check of the formula $\inferredSpecPhysical{1} \Rightarrow \specPhysical{1}$ is valid, that indicates $\inferredSpecPhysical{1} $ is a candidate invariant of the AFC system. Next, we vary the input values and observe the consequent behaviors of the system.

First, we vary the value of the engine speed and keep other parameters unchanged. Assuming $w = 2200rpm$, the inferred physical specification of the air-fuel ratio from \toolhynger and Daikon becomes $\inferredSpecPhysical{1}  \deq \ mode = startup \wedge t \leq 9.5 \Rightarrow 14.129 \leq \Vout(t) \leq 15.033$.
Hence, the formula $\inferredSpecPhysical{1}  \Rightarrow \specPhysical{1}$ is false indicating that the AFC system may contain a cyber-physical specification mismatch as we change the engine speed input.

Second, we change the range of the throttle input to $[40^{\circ}, 70^{\circ}]$. Then, the inferred physical specification of the air-fuel ratio from \toolhynger and Daikon becomes $\inferredSpecPhysical{1}  \deq \ mode = startup \wedge t \leq 9.5 \Rightarrow 14.396 \leq \Vout(t) \leq 14.849$.
Hence, $\inferredSpecPhysical{1} $ no longer implies $\specPhysical{1}$. Therefore, there exists a cyber-physical specification mismatch when changing the throttle input as well.

\paragraph*{Cyber-physical specification mismatches according to $\specPhysical{2}$}
Initially, the physical specification in \eqref{fuel_normal} is $\specPhysical{2} \deq \ mode = normal \wedge t \geq 9.5 \Rightarrow 14.406 \leq \lambda(t) \leq 14.994$.
Here, the magnitude bound of the air-fuel ratio at the $normal$ mode of the system inferred from \toolhynger and Daikon is $\inferredSpecPhysical{2}  \deq \ mode = normal \wedge t \geq 9.5 \Rightarrow 14.645 \leq \lambda(t) \leq 14.84$.
Then, we can consider $\inferredSpecPhysical{2}$ as a candidate invariant of the system because the formula $\hat{\varphi}_P \Rightarrow \specPhysicalSymbol$ is true.
%

Next, we investigate whether there is a specification mismatch for the AFC system as we change the proportional and integral gains of its PI controller.
\tabref{fuel_PI} describes the comparison between the actual physical specification $\specPhysical{2}$ and the physical specification $\inferredSpecPhysical{2}$ inferred from \toolhynger and Daikon, where $\inferredSpecPhysical{2} \downarrow \lambda$ denotes the inferred bound for $\lambda$ when $t \geq t_s$ and $mode = normal$.
In \tabref{fuel_PI}, the check of the formula $\inferredSpecPhysical{2} \Rightarrow \specPhysical{2}$ returns false in some cases (e.g., when $c_{13} = 0.04$, $c_{14} = 0.04$) indicating that the changes in the controller gains may produce cyber-physical specification mismatches for the AFC system.
\begin{table*}[t!]
	\centering
	\small{
		\begin{tabular}{|c|c|c|c|}
		\hline
		Controller Gain & $\inferredSpecPhysical{2} \downarrow \lambda$ & $\inferredSpecPhysical{2}  \Rightarrow \specPhysical{2}$ & $\specPhysical{2} \Rightarrow \inferredSpecPhysical{2}$ \\ \hline
		$c_{13} = 0.01$, $c_{14} = 0.14$ &$14.567 \leq \lambda(t) \leq 15.058$&$False$&$False$ \\ \hline
		$c_{13} = 0.02$, $c_{14} = 0.14$ &$14.592 \leq \lambda(t) \leq 15.033$&$False$&$False$ \\ \hline
		$c_{13} = 0.06$, $c_{14} = 0.14$ &$14.634 \leq \lambda(t) \leq 14.955$&$True$&$False$ \\ \hline
		$c_{13} = 0.8$, $c_{14} = 0.14$ &$14.642 \leq \lambda(t) \leq 14.929$&$True$&$False$ \\ \hline
		$c_{13} = 0.04$, $c_{14} = 0.04$ &$14.649 \leq \lambda(t) \leq 15.007$&$False$&$False$ \\ \hline
		$c_{13} = 0.04$, $c_{14} = 0.34$ &$14.581 \leq \lambda(t) \leq 14.937$&$True$&$False$ \\ \hline
		$c_{13} = 0.04$, $c_{14} = 0.64$ &$14.577 \leq \lambda(t) \leq 14.888$&$True$&$False$ \\ \hline
		$c_{13} = 0.04$, $c_{14} = 0.94$ &$14.589 \leq \lambda(t) \leq 14.855$&$True$&$False$ \\ \hline
		\end{tabular}}
		\caption{Experiment results illustrate the comparison between actual physical specifications and inferred physical invariants from \toolhynger and Daikon of the AFC system when changing the proportional gain and the integral gain of its PI controller.}
	\tablabel{fuel_PI}%
	\vspace{-2em}
\end{table*}

%% file: discussion.tex
\section{Discussion}
\seclabel{discussion}
Identifying a cyber-physical specification mismatch of CPS with dynamic analysis is a challenging problem. Although the \toolhynger prototype in conjunction with Daikon can detect potential cyber-physical specification mismatches of CPS, such as those in the case studies described in \secref{experiment}, however, it has some limitations.
%
%
First, the Daikon tool used by \toolhynger may only infer extremely limited classes of nonlinear invariants by default (e.g., squares like $x^2$), and not general polynomials (e.g., $x^2+y^2+z^3$). So we plan to extend the invariant templates to be able to capture more interesting relations, particularly for physical variables.
Second, although Daikon can infer candidate invariants in terms of logical predicates over variables, it has limitation for checking complex specifications related to real-time requirements such as STL, MTL and HyperSTL~\cite{nguyen2017memocode}.
Industrial-scale CPS usually have safety and liveness requirements depending on precise real-time relations of signals, so strengthening the capability of checking temporal logic like STL, MTL and HyperSTL in Daikon would leverage the methodology presented in this paper.
%
%

%
Additionally, while the \toolhynger tool is a prototype, it can be envisioned to take an arbitrary SLSF model, instrument it, feed the resulting traces to Daikon to generate candidate invariants, then check if these candidate invariants are actual invariants or not (using, e.g., SpaceEx~\cite{frehse2011cav} or other hybrid system model checkers), as well as identify specification mismatches.
For example, the candidate invariants inferred from \toolhynger and Daikon of the buck converter including only plant and controller represented in term of hybrid automata in~\figref{automaton_buckboost} would easily be checked to see whether they are actually invariants using SpaceEx.
In long term, \toolhynger could be extended for runtime assurance tasks like detecting and thwarting security violations and attacks, similar to the ClearView tool that also uses Daikon~\cite{perkins2009sosp}.
ClearView's success for software systems illustrates that finding sets of candidate invariants and monitoring their evolution over time may be useful for runtime assurance and resiliency methods in CPS.
If the candidate invariants are checked at runtime using a real-time reachability method~\cite{bak2014rtss}, a formal and dynamic runtime assurance environment may be feasible.
%

%% file: conclusion.tex
\section{Conclusion \& Future Works}
\seclabel{conclusion}
The results illustrate the feasibility of using dynamic invariant inference for analysis of embedded and cyber-physical systems.
The \toolhynger prototype enables a powerful extension of dynamic invariant inference to CPS for two main reasons.
First, it enables potentially model-free and black box invariant inference, since the internals of the SLSF blocks may remain unknown.
If no model is available (in the black box case), the candidate invariants represent what may be the most formal model available, otherwise (in the white box case), then candidate invariants represent a candidate abstraction of that model. If the candidate invariants are actual invariants, this is powerful, as they represent what is likely a less complex representation of the set of reachable states of the system.
%
%
Second, if we view the SLSF models as hybrid automata in a formal context, it represents the first use of dynamic execution analysis for hybrid systems with sophisticated software state and discrete complexity. Two proof-of-concept CPS case studies including the DC-to-DC power converter and the powertrain fuel control system are presented to illustrate the capability of \toolhynger in detecting potential cyber-physical specification mismatches.

Overall, there are several directions for future research, including:
\begin{enumerate*}[label=(\textit{\alph*})]
\setlength\itemsep{0.1em}
\item extending the classes of invariants that may be inferred, particularly to nonlinear (polynomial)~\cite{nguyen2012icse} and disjunctive/max-plus forms~\cite{nguyen2014icse}, potentially by integrating Daikon with techniques from Dig~\cite{nguyen2014tosem},
%
%
\item runtime assurance and verification with real-time reachability of inferred invariants~\cite{bak2014rtss},
\item improving and refining \toolhynger, particularly with regard to performance (such as using Daikon in the online mode with direct pipes between \toolhynger and Daikon, so that file I/O is minimized), and
\item analyzing more industrial-scale CPS using \toolhynger.
%
\end{enumerate*} 